\documentclass[pra,letterpaper,aps,10pt,superscriptaddress,twocolumn,floatfix,showpacs]{revtex4-1}

\usepackage{amsmath, upgreek, amssymb, graphicx, subfigure, paralist}
\usepackage[colorlinks, linkcolor=blue, citecolor=blue, urlcolor=blue, breaklinks=true]{hyperref}

\newcommand{\eref}[1]{Eq.~(\ref{#1})}

\newcommand{\fref}[1]{Fig.~\ref{#1}}
\newcommand{\frefs}[1]{Figs.~\ref{#1}}
\newcommand{\Fref}[1]{Figure~\ref{#1}}

\newcommand{\sref}[1]{Sec.~\ref{#1}}
\newcommand{\Sref}[1]{Section~\ref{#1}}

\newcommand{\rmd}{\text{d}}

\newcommand{\twovec}[2]{\begin{pmatrix}#1\\#2\end{pmatrix}}
\newcommand{\twotwomat}[4]{\begin{bmatrix}#1&#2\\#3&#4\end{bmatrix}}
\newcommand{\abs}[1]{\lvert#1\rvert}

\newcommand{\im}[1]{\text{Im}\!\left\{#1\right\}}
\newcommand{\re}[1]{\text{Re}\!\left\{#1\right\}}

\DeclareMathOperator{\arccot}{arccot}

\makeatletter
\newlength \figurewidth
\makeatother

\begin{document}


\title{Collectively-enhanced optomechanical coupling in periodic arrays of scatterers}

\date{\today}

\author{Andr\'e Xuereb}
\email[Corresponding author. ]{andre.xuereb@um.edu.mt}
\affiliation{Department of Physics, University of Malta, Msida MSD\,2080, Malta}
\affiliation{Centre for Theoretical Atomic, Molecular and Optical Physics, School of Mathematics and Physics, Queen's University Belfast, Belfast BT7\,1NN, United Kingdom}
\author{Claudiu Genes}
\affiliation{Institut f\"ur Theoretische Physik, Universit\"at Innsbruck, Technikerstrasse 25, A-6020 Innsbruck, Austria}
\affiliation{ISIS (UMR 7006) and IPCMS (UMR 7504), Universit\'{e} de Strasbourg and CNRS, Strasbourg, France}
\author{Aur\'elien Dantan}
\affiliation{QUANTOP, Danish National Research Foundation Center for Quantum Optics, Department of Physics and Astronomy, University of Aarhus, 8000 Aarhus C, Denmark}

\begin{abstract}
We investigate the optomechanical properties of a periodic array of identical scatterers placed inside an optical cavity and extend the results of [A.\ Xuereb, C.\ Genes, and A.\ Dantan, Phys.\ Rev.\ Lett.\ \textbf{109}, 223601 (2012)]. We show that operating at the points where the array is transmissive results in linear optomechanical coupling strengths between the cavity field and collective motional modes of the array that may be several orders of magnitude larger than is possible with an equivalent reflective ensemble. We describe and interpret these effects in detail and investigate the nature of the scaling laws of the coupling strengths for the different transmissive points in various regimes.
\end{abstract}

\maketitle

The ability to measure and control the motion of massive mechanical oscillators has progressed dramatically in recent years~\cite{Aspelmeyer2013,Meystre2013}, and several important milestones have been reached in the field of optomechanics towards bringing this capability into the quantum regime, including the cooling to the motional quantum ground state~\cite{Teufel2011,Chan2011,Verhagen2012}, the detection of quantized mechanical motion~\cite{SafaviNaeini2012,Brahms2012}, and the observation of the ponderomotive squeezing of light~\cite{Brooks2012,SafaviNaeini2013} or of the radiation-pressure shot noise on a mechanical oscillator~\cite{Purdy2013}.

One challenge faced by the current generation of optomechanical experiments is that the interaction strength between a single photon and a single massive mechanical element is typically very weak. This can be ameliorated by confining light in wavelength-scale structures~\cite{Eichenfield2009} or, generically, counteracted by the use of strong light fields in an optical resonator to amplify the interaction strength~\cite{Groblacher2009a}, albeit at the expense of trading off the intrinsically nonlinear nature of the radiation-pressure interaction (see, however, the recent proposals in Refs.~\cite{Borkje2013,Lemonde2013}). A growing number of theoretical proposals has contemplated the opposite -- `strong coupling' -- regime, where a single photon can affect the motion of the oscillator significantly, thus giving access to the full quantum nature of the optomechanical interaction~\cite{Akram2010,Rabl2011,Nunnenkamp2011,Nunnenkamp2012,Qian2012,Ludwig2012,Stannigel2012,Kronwald2013,Liao2013}.

On the other hand, \emph{collective} effects in optomechanical systems involving multiple mechanical and electromagnetic field modes have been discussed in a number of theoretical works, in connection with, e.g., optomechanical entanglement~\cite{Mancini2002,Zhang2003,Marshall2003,Eisert2004,Pinard2005,Vitali2007b,Genes2008d,Bhattacharya2008,Bhattacharya2008a,Hartmann2008,Ludwig2010}, enhanced displacement sensitivity~\cite{Miao2009,Dobrindt2010,Xu2013}, optomechanical nonlinearities~\cite{Stannigel2012,Ludwig2012,Seok2012,Komar2013,Genes2013}, quantum information processing~\cite{Stannigel2010,Chang2011,Jacobs2012,Rips2012}, many-body physics~\cite{Heinrich2010,Heinrich2011,Marcos2012,Akram2012,Ludwig2012b}, as well as in a number of recent experiments~\cite{Grudinin2010,Lin2010,Mahboob2011,Mahboob2012,massel2011,Zhang2012,Botter2013}. Collective optomechanical effects are also at the heart of cavity optomechanics with cold atomic ensembles~\cite{StamperKurn2012}.

Motivated by the exploration of such collective optomechanical effects, we recently~\cite{Xuereb2012c} showed that the collective motion of a periodic array of identical scatterers, when placed inside a cavity field, can couple very strongly to the optical field in the configuration where the array is \emph{transmissive}, in contrast to the usual \emph{reflective} optomechanics approach. The aim of the present work is to present a detailed exploration of this system, to highlight the regimes in which these generic collective effects are seen and to compare the various possible transmissive operating points.
\par
This paper is organized as follows. In the next section we summarize and discuss the tools used to model a periodic array of $N$ identical elements, and show that such an array can be modeled as a single effective element in the framework of the transfer matrix theory for one-dimensional scatterers. \Sref{sec:OM} discusses the optomechanical properties of such a generic $N$-element stack, when placed inside an optical cavity, in two distinct and opposite regimes: (i)~A maximally-reflective stack (\sref{sec:OM:Refl}), and (ii)~a transmissive stack (\sref{sec:OM:Tran}). In the second regime we show that the equations of motion for the system at hand are effectively described by an optical field interacting with a single collective mechanical mode, whose profile strongly depends on the transmissive operating point chosen. The next section discusses this regime in further detail, and explores the scaling of the optomechanical coupling strength in the transmissive regime with the various system parameters and the various operating points; we show that the increase in the optomechanical coupling strength with the number of scatterers combines with an effect whereby the linewidth of the cavity resonance is narrowed due to the presence of the stack to provide an enhancement of the optomechanical cooperativity by several orders of magnitude over that of a single element system. A detailed comparison of the different transmissive points of the system is also given. Finally, \Sref{sec:Imp} briefly examines the case of absorbing elements.

\begin{figure}[t]%
 \includegraphics{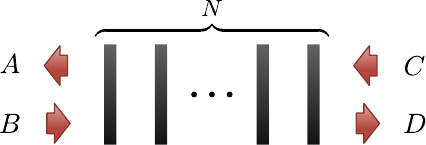}%
\caption{Schematic diagram illustrating the basis of the model: $N$ scatterers interacting with four running waves. $B$ and $C$ represent incident fields, $A$ and $D$ outgoing fields. The four field amplitudes are connected by means of transfer matrices.}%
 \label{fig:Stack}%
\end{figure}%
\section{Modeling an $N$-element stack}
Throughout this paper we shall restrict ourselves to a one-dimensional system and use the transfer matrix formalism to model a periodic $N$-element array, as illustrated in \fref{fig:Stack}. As is well-known~\cite{Deutsch1995,Xuereb2009b}, this formalism can treat elements (`scatterers') that interact linearly with the electromagnetic field, no matter the strength of this interaction or whether the scatterer is lossless or not. Within this formalism, each scatterer is parametrized by its polarizability $\zeta$, which is real for lossless scatterers but complex in the presence of absorption. $\zeta$ is related to the amplitude reflectivity $r$ of the element through the expression
\begin{equation}
r=\frac{i\zeta}{1-i\zeta}\,.
\end{equation}
Therefore, for a lossless scatterer, we have
\begin{equation}
\abs{r}^2=\frac{\zeta^2}{1+\zeta^2}\,,
\end{equation}
which allows us to link our results to, e.g., those in Ref.~\cite{Miladinovic2011} through their Eq.\ (8). The flexibility afforded by the transfer matrix formalism allows us to treat ensembles of atoms in an optical lattice on the same footing as periodic arrays of macroscopic scattering elements (e.g., arrays of thin dielectric membranes) in the limit of a one-dimensional scattering theory. The formal relation between these physical systems and the transfer matrix formalism, through the parameter $\zeta$, was illustrated in the Supplemental Information for Ref.\ \cite{Xuereb2012c}.

\subsection{Transfer matrix for $N$-element stack}
We start by discussing the optical properties of a periodic array of $N$ equally-spaced elements, assumed to be identical to one another and non-absorbing for the time being. Each element is assumed to have a small thickness compared to the wavelength of the light in question, as explained in further detail in the Supplemental Information for Ref.\ \cite{Xuereb2012c}. At this point we make no distinction between arrays of membranes and arrays of atoms in an optical lattice; the following formalism holds identically for either case.\\
The optical properties of the array will be determined entirely by the number of elements, the distance $d$ between pairs of elements in the array, and the polarizability $\zeta$ of each element. A real $\zeta$ captures the fact that there is no absorption in the elements; this requirement will be lifted later on. As a first step, we recall the matrix relating the electromagnetic fields interacting with a single element~\cite{Deutsch1995},
\begin{equation}
M_\mathrm{m}(\zeta)\equiv\twotwomat{1+i\zeta}{i\zeta}{-i\zeta}{1-i\zeta}\,.
\end{equation}
To describe our array we also need the effect of free-space propagation of a monochromatic beam of wavelength $\lambda=2\pi/k$ over a distance $d$,
\begin{equation}
M_\mathrm{p}(d)\equiv\twotwomat{e^{ikd}}{0}{0}{e^{-ikd}}\,.
\end{equation}
Crucially, both $M_\mathrm{m}(\zeta)$ and $M_\mathrm{p}(d)$ have unit determinant. These matrices relate forward- and backward-propagating electromagnetic waves on either side of the element:
\begin{equation}
\twovec{A}{B}=M\cdot\twovec{C}{D}\,,
\end{equation}
with $A$ and $C$ being the (complex) amplitudes of the backward-propagating waves, and similarly $B$ and $D$ the amplitudes of the forward-propagating waves. \Fref{fig:Stack} illustrates the situation we wish to describe. The transfer matrix of the array can be written as a product of the form
\begin{equation}
M_\mathrm{m}(\zeta)\cdot M_\mathrm{p}(d)\cdot M_\mathrm{m}(\zeta)\cdots M_\mathrm{m}(\zeta)\,,
\end{equation}
where $M_\mathrm{m}(\zeta)$ appears $N$ times (i.e., once for each element). We now define a matrix $M$ such that
\begin{multline}
M_\mathrm{p}(d/2)\cdot M_\mathrm{m}(\zeta)\cdot M_\mathrm{p}(d)\cdots M_\mathrm{m}(\zeta)\cdot M_\mathrm{p}(d/2)\\
=\bigl[M_\mathrm{p}(d/2)\cdot M_\mathrm{m}(\zeta)\cdot M_\mathrm{p}(d/2)\bigr]^N\equiv M^N\,.
\end{multline}
Evaluating the product explicitly, $M$ can be written as
\begin{equation}
M\equiv\twotwomat{(1+i\zeta)e^{ikd}}{i\zeta}{-i\zeta}{(1-i\zeta)e^{-ikd}}\,.
\end{equation}
Once again, it is apparent that $M$ has unit determinant. This property is crucial, for it allows us to write~\cite{Yeh2005}, for the case of real $\zeta$,
\begin{equation}
M^N=\twotwomat{(1+i\chi)e^{i(kd+\mu)}}{i\chi}{-i\chi}{(1-i\chi)e^{-i(kd+\mu)}}\,,
\end{equation}
where $\chi\equiv\zeta U_{N-1}(a)$, with $U_n(x)$ being the $n$\textsuperscript{th} Chebyshev polynomial of the second kind, $a=\cos(kd)-\zeta\sin(kd)$, and
\begin{equation}
e^{i\mu}=\frac{1-i\zeta
U_{N-1}(a)}{(1-i\zeta)U_{N-1}(a)-e^{ikd}U_{N-2}(a)}\,.
\end{equation}
The matrix $M^N$ has an extra `padding' of $d/2$ on either side. We remove this padding to obtain, finally, the transfer matrix describing the $N$-element array:
\begin{equation}
M_N\equiv M_\mathrm{p}[\mu/(2k)]\cdot M_\mathrm{m}(\chi)\cdot M_\mathrm{p}[\mu/(2k)]\,.
\end{equation}
What is remarkable about $M_N$ is that the $N$ (lossless) elements behave as a single collective `superelement' of polarizability $\chi$, supplemented with a `padding' equivalent to a phase-shift of $\mu/2$ on either side of the array. This fact not only aids interpretation of the optical properties of the stack, but also simplifies the algebra involved considerably.\par
Generically, the transfer matrix of any optical system can be related to the (amplitude) transmissivity and reflectivity of that same system. For concreteness, let us suppose that the system at hand can be described by a transfer matrix of the form
\begin{equation}
\twotwomat{m_{11}}{m_{12}}{m_{21}}{m_{22}}\,,
\end{equation}
where the four entries are determined by the optical properties of the system at hand. The complex transmissivity of the system can then be written down as
\begin{equation}
\label{eq:TransGeneral}
\mathcal{T}=\frac{1}{m_{22}}\,.
\end{equation}
Correspondingly, its reflectivity is
\begin{equation}
\mathcal{R}=\frac{m_{12}}{m_{22}}\,,
\end{equation}
\begin{figure}[t]%
 \includegraphics[width=\figurewidth]{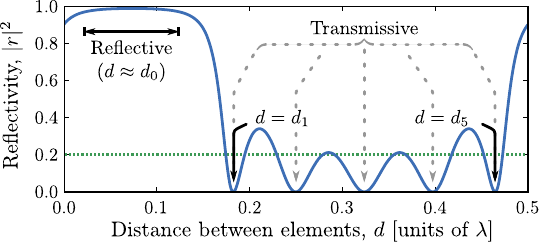}%
\caption{Optical properties of a stack of $N=6$ non-absorbing elements. The intensity reflectivity of the stack (blue curve) varies strongly, from $0$\% to about $99$\%, as the distance between the elements is scanned; this curve is periodic with a period of $\lambda/2$. By way of comparison, we show the corresponding reflectivity for a single element as the dotted green line. We mark three important values of $d$ in this figure:\ $d=d_1$ and $d=d_5$, where the reflectivity is zero, and $d=d_0$, where the reflectivity attains its largest value.}%
 \label{fig:Reflectivity}%
\end{figure}%
These quantities are used throughout this paper to characterize the optical properties of our system. A plot of the intensity reflectivity of a lossless $6$-element ensemble as the spacing between pairs of elements is varied is shown in \fref{fig:Reflectivity}. Note that, despite each element having a reflectivity of $20$\%, the reflectivity of the entire ensemble ranges from $0$\% (e.g., at $d=d_1$) to about $99$\% (at $d=d_0$). Moreover, an ensemble of $N$ elements possesses $N-1$ points $d_l$ ($l=1,\dots,N-1$) at which the transmission is zero; in the figure we label only the outermost two of these points, $d_1$ and $d_5$. Unlike a standard distributed Bragg reflector, rather than choosing the ratio of the refractive indices between the two dielectrics making up the structure (one of the dielectrics being vacuum in our case), we control the optical path length of the vacuum layers through $d$. In our structure, therefore, the thickness of the elements is completely decoupled from the value of $d$.

\section{Optomechanics of a periodic array of scatterers}\label{sec:OM}
Our next task is to place the array just described inside a near-resonant optical cavity. The interaction of the array with the cavity field will shift the resonances of the main cavity. As is usual in optomechanics, we assume that each element in the array is harmonically trapped, and are interested in one particular figure of merit:\ the coupling strength defined as the frequency shift incurred by the cavity resonance when the array undergoes a displacement equal to the size of the relevant zero-point-fluctuations. As a yardstick we shall use the quantity
\begin{equation}
g\equiv\frac{2\omega_\mathrm{c}x_\mathrm{zpt}}{L}\,,
\end{equation}
which is the optomechanical coupling strength for a perfectly reflective mirror near the center of a cavity of length $L$ and resonant frequency $\omega_\mathrm{c}$~\cite{Xuereb2012d}. The size of the zero-point-fluctuations of each element is denoted by $x_\mathrm{zpt}=\sqrt{\hbar/(m\omega_\mathrm{m})}$, where $m$ is the effective mass of the element and $\omega_\mathrm{m}$ its oscillation frequency.

\subsection{Element stack inside a cavity}
\begin{figure}[t]
 \includegraphics[width=\figurewidth]{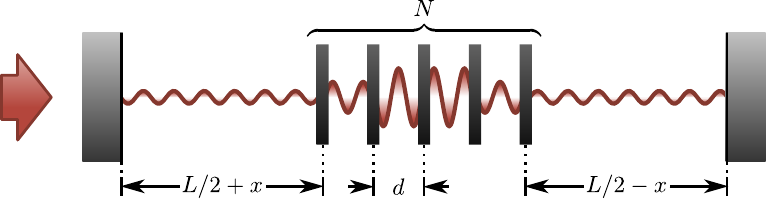}
\caption{A schematic drawing of the generic system we are considering. A periodic array of $N$ elements, each of which is independently harmonically-bound, is positioned at, or very close to, the center of a Fabry--P\'erot resonator. Throughout this paper, we shall consider only the case for which $L\gg Nd$.}
 \label{fig:Cavity}
\end{figure}

\Fref{fig:Cavity} illustrates schematically the periodic element array placed inside a Fabry--P\'erot cavity of length $L$, assumed much longer than the array ($L\gg Nd$). The transfer matrix describing this system is then
\begin{multline}
M_\mathrm{cav}\equiv M_\mathrm{m}(Z)\cdot M_\mathrm{p}(L/2+x)\\
\cdot M_N\cdot M_\mathrm{p}(L/2-x)\cdot M_\mathrm{m}(Z)\,.
\end{multline}
Here $x$ is the displacement of the ensemble with respect to its position at the center of the cavity and $Z$ is the polarizability of the cavity mirrors, assumed equal for both. For good, lossless, cavity mirrors ($\abs{Z}\gtrsim1$), the finesse of the cavity may be simply written $\mathcal{F}=\pi\abs{Z}\sqrt{Z^2+1}$~\cite{Xuereb2012a}. The transmission of the system, following \eref{eq:TransGeneral}, is given by
\begin{equation}
\mathcal{T}_\mathrm{cav}=\frac{1}{\bigl(M_\mathrm{cav}\bigr)_{22}}\,;
\end{equation}
the maxima of $\mathcal{T}_\mathrm{cav}$ give the resonances of this system. In order to find these resonances analytically, we consider a simpler system where the cavity mirrors are perfect; we need only solve the relation
\begin{multline}
\twovec{\phantom{+}1}{-1}\varpropto\twotwomat{e^{i\theta}}{0}{0}{e^{-i\theta}}\times\twotwomat{1+i\chi}{i\chi}{-i\chi}{1-i\chi}\\
\times\twotwomat{e^{i\phi}}{0}{0}{e^{-i\phi}}\cdot\twovec{\phantom{+}1}{-1}\,,
\end{multline}
with $\theta\equiv k(L/2+x)+\mu/2$ and $\phi\equiv k(L/2-x)+\mu/2$. We thus obtain
\begin{equation}
\label{eq:ResonanceSolution}
e^{ikL}=\frac{e^{-i\mu}}{1+i\chi}\biggl[i\chi\cos(2kx)\pm\sqrt{1+\chi^2\sin^2(2kx)}\biggr]
\end{equation}
However, we immediately see that this equation is transcendental in $k$, and therefore cannot be solved analytically; this equation is easily solvable for $L$, however, given a fixed operating wavelength.

\subsection{Center-of-mass coupling: Reflective optomechanics}\label{sec:OM:Refl}
It is now a legitimate question to ask: `If $d$ (or $x$) shifts by a small amount, how much will the resonant frequency of \emph{this} cavity shift?' This question is easily answered by expanding \eref{eq:ResonanceSolution} in small increments about its solution. Assuming a dominantly linear effect, we replace $k\to k+\delta k$, $x\to x+\delta x$, $\chi\to\chi+\delta\chi$, and $\mu\to\mu+\delta\mu$ in \eref{eq:ResonanceSolution}. Around resonance, the result simplifies to
\begin{align}
\label{eq:GeneratorEquation}
L\delta k+\delta\mu=&\ \biggl[-1\pm\cos(2kx)\Big/\sqrt{1+\chi^2\sin^2(2kx)}\biggr]\nonumber\\
&\qquad\qquad\times\delta\chi/\bigl(1+\chi^2\bigr)\nonumber\\
&\mp\biggl[2\chi\sin(2kx)\Big/\sqrt{1+\chi^2\sin^2(2kx)}\biggr]\nonumber\\
&\qquad\qquad\times(x\delta k+k\delta x)\,.
\end{align}
For the rest of this section, we shall consider the center-of-mass motion of the ensemble, and use \eref{eq:GeneratorEquation} to compute the optomechanical coupling strength. For such a uniform displacement, $\partial\mu=\partial\chi=0$, and we assume that $\abs{L/x}$ is very large, such that we can write
\begin{equation}
L\delta k=\mp\biggl[2\chi\sin(2kx)\Big/\sqrt{1+\chi^2\sin^2(2kx)}\biggr]k\delta x\,.
\end{equation}
The right-hand-side of this equation is maximized when $\sin(2kx)=\mp1$, whereby
\begin{equation}
L\delta k=2k\bigl(-\chi\big/\sqrt{1+\chi^2}\bigr)\delta x\,.
\end{equation}
\begin{figure}[t]%
 \includegraphics[width=\figurewidth]{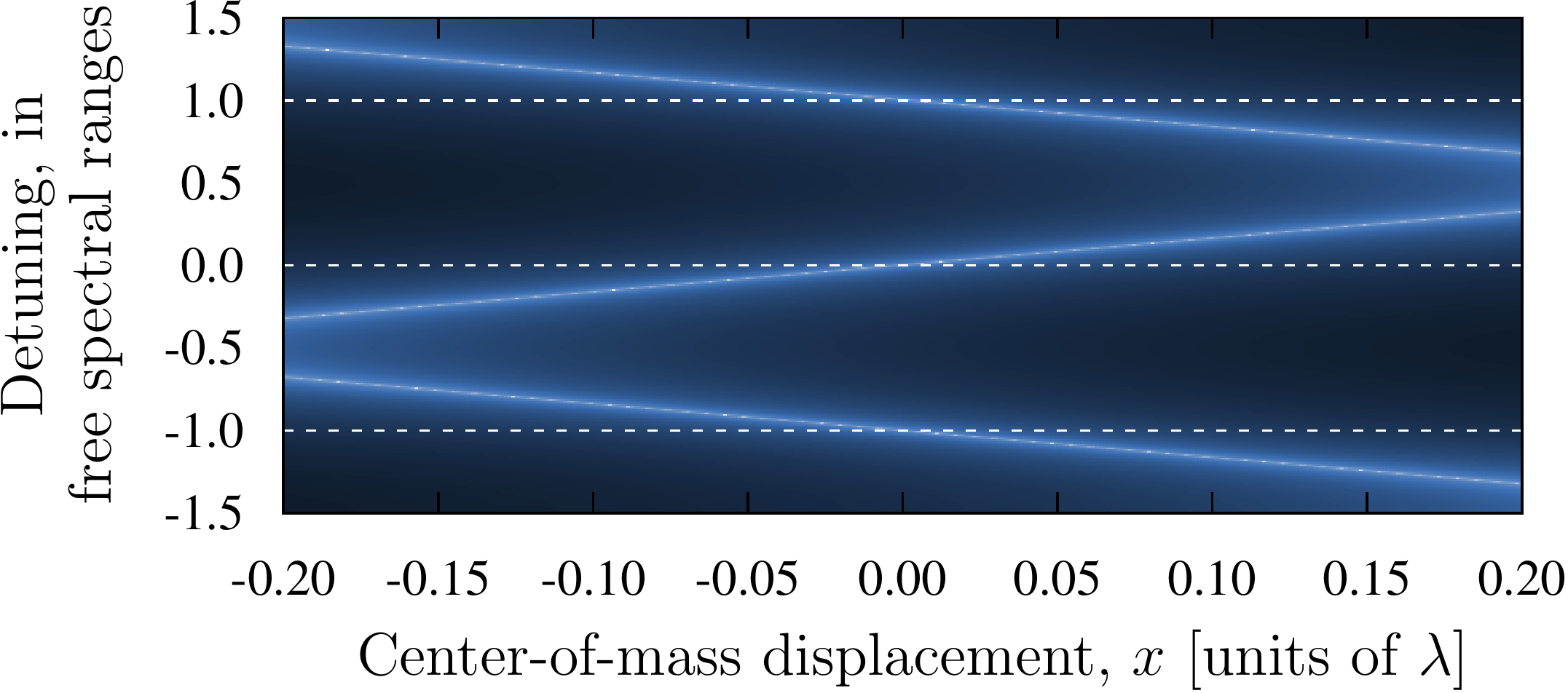}%
\caption{Transmission through a cavity with $N=6$ immobile elements configured for maximal reflectivity. The dashed lines denote the bare-cavity resonances, which are shifted due to the presence of the ensemble. $x$ is normalized by a factor $\sqrt{N}$, such that the gradient of the bright curves gives directly the linear optomechanical coupling at that point. ($\zeta=-0.5$, $L\approx6.3\times10^4\lambda$, $d=d_1$, bare-cavity finesse\ $\approx\,3\times10^4$, corresponding to cavity-mirror reflectivities of $99.99$\%.}%
 \label{fig:TransmissionCOM}%
\end{figure}%
This is, in absolute value, a monotonically-increasing function of $\abs{\chi}$ and is therefore maximized when $\chi$ attains its largest value, $\chi_0\equiv\zeta\,U_{N-1}\bigl(\sqrt{1+\zeta^2}\bigr)=-i\sin\bigl[N\cos^{-1}\bigl(\sqrt{1+\zeta^2}\big)\bigr]$. The ensemble attains this reflectivity for when $kd=kd_0\equiv-\tan^{-1}(\zeta)$ (see \fref{fig:Reflectivity}). The resulting coupling strength can then be shown to be
\begin{equation}
g_\mathrm{com}=g\sqrt{\mathcal{R}/N}\,,
\end{equation}
where $\mathcal{R}=\chi_0^2\big/\bigl(1+\chi_0^2\bigr)$. The factor of $1/\sqrt{N}$ that is introduced into this expression has a natural explanation: The motional mass of $N$ elements is $Nm$, and therefore the scale of the zero-point-fluctuations is $x_\mathrm{zpt}/\sqrt{N}$. We can draw two immediate conclusions regarding $g_\mathrm{com}$: (i)~$g_\mathrm{com}\leq g$, and (ii)~$g_\mathrm{com}$ is optimized for $\mathcal{R}\to1$. A plot of the transmission spectrum of a cavity with a $6$-element ensemble inside it is shown in \fref{fig:TransmissionCOM} as the position of the ensemble is varied. The gradient of the bright curves at each point gives directly the linear optomechanical coupling strength at that point.\\
The essential aim of this paper is to outline a mechanism~\cite{Xuereb2012c} whereby coupling strengths much larger than $g$ can be obtained, despite keeping $\omega_\mathrm{c}$ and $L$ fixed. To do this, we shall now explore the coupling strength to the motion of \emph{individual} elements, rather than to the ensemble as a whole.

\subsection{Coupling to each individual element in the transmissive regime}\label{sec:OM:Tran}
Just as the ensemble attains its peak reflectivity for $d=d_0$, we can see from \fref{fig:Reflectivity} that its reflectivity is zero at $d=d_{l}$, where each $d_l$ (for real $\zeta$), for $l=1,\dots,N-1$ is defined, modulo $\lambda/2$, by
\begin{equation}
d_l\equiv\frac{1}{k}\biggl\{\cos^{-1}\Bigl[\cos(l\pi/N)/\sqrt{1+\zeta^2}\Bigr]-\tan^{-1}(\zeta)\biggr\}\,.
\end{equation}
We now work with one such inter-element separation and obtain the optomechanical coupling strength, i.e., the shift in cavity resonance frequency due to the motion of each element in the ensemble. To allow one element, say the $j$\textsuperscript{th}, to move independently of the rest of the ensemble, we conceptually split the ensemble into three sections:\ the elements to the `left' of $j$, the $j$\textsuperscript{th} element itself, and the elements to the `right' of $j$. With this logic, the matrix $M_N$ representing the ensemble can be written, for $1\leq j\leq N$,
\begin{multline}
\twotwomat{e^{i\mu_1/2}}{0}{0}{e^{-i\mu_1/2}}\twotwomat{1+i\chi_1}{i\chi_1}{-i\chi_1}{1-i\chi_1}\\
\times\twotwomat{e^{i(\mu_1/2+\nu+k\delta x_j)}}{0}{0}{e^{-i(\mu_1/2+\nu+k\delta x_j)}}\\
\times\twotwomat{1+i\zeta}{i\zeta}{-i\zeta}{1-i\zeta}\twotwomat{e^{i(\mu_2/2+\nu-k\delta x_j)}}{0}{0}{e^{-i(\mu_2/2+\nu-k\delta x_j)}}\\
\times\twotwomat{1+i\chi_2}{i\chi_2}{-i\chi_2}{1-i\chi_2}\twotwomat{e^{i\mu_2/2}}{0}{0}{e^{-i\mu_2/2}}\,,
\end{multline}
where $\nu=kd_l$, $\mu_1$ and $\chi_1$ describe the ensemble formed by the $n_1=j-1$ membranes to the `left' of the $j$\textsuperscript{th}, and $\mu_2$ and $\chi_2$ the one formed by the $n_2=N-j$ membranes to its `right'. The displacement of the $j$\textsuperscript{th} element is denoted $\delta x_j$; all other membranes are in their equilibrium position. This small displacement shifts the resonance frequency of the cavity $\omega\to\omega-g_j^{(l)}\delta x_j$, defining $g_j^{(l)}$ as the optomechanical coupling strength for the $j$\textsuperscript{th} element when $d=d_l$. In the transmissive regime, to lowest order in $k\delta x_j$ in each entry, the matrix product above can be written, with the above choice for $\nu$,
\begin{equation}
\twotwomat{e^{i\mu}+\alpha\,\delta x_j}{\beta\,\delta x_j}{\beta^\ast\,\delta x_j}{e^{-i\mu}+\alpha^\ast\,\delta x_j}\,,
\end{equation}
where $\alpha$ and $\beta$ are increments of first order in the relevant displacement [note that the (off-)diagonal terms are complex conjugates of each other; this is different to the case where absorption is nonzero]. When this matrix is substituted into the equation for the resonance condition, the terms involving $\re{e^{-i\mu}\alpha}$ and $\re{\beta}$ drop out entirely \emph{for a symmetric system}, such that it suffices to consider only the imaginary part of the increment. Let us reiterate that this happens only because the off-diagonal terms are complex conjugates of each other; were absorption to be nonzero, this would no longer be the case. \eref{eq:GeneratorEquation} now simplifies to
\begin{figure}[t]%
 \includegraphics[width=\figurewidth]{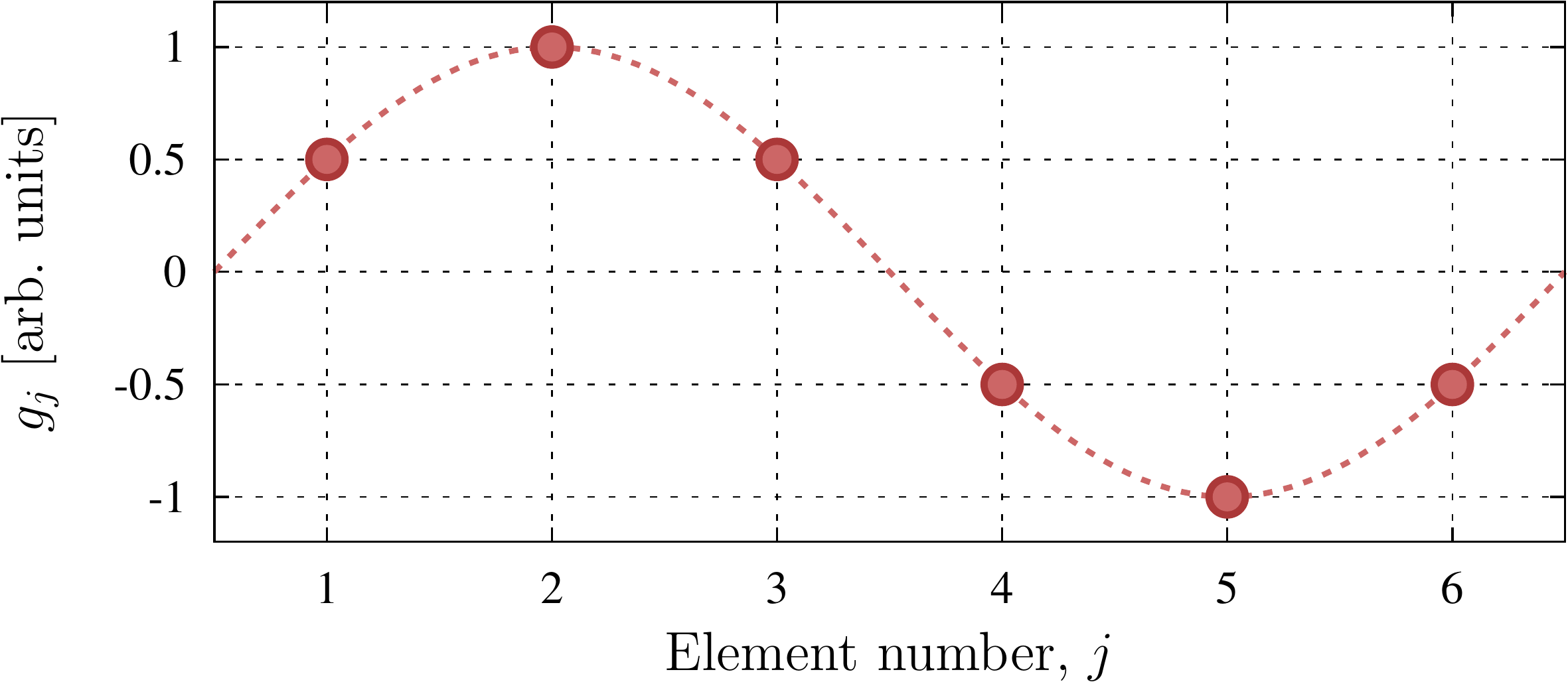}\\[2mm]%
 \includegraphics[width=\figurewidth]{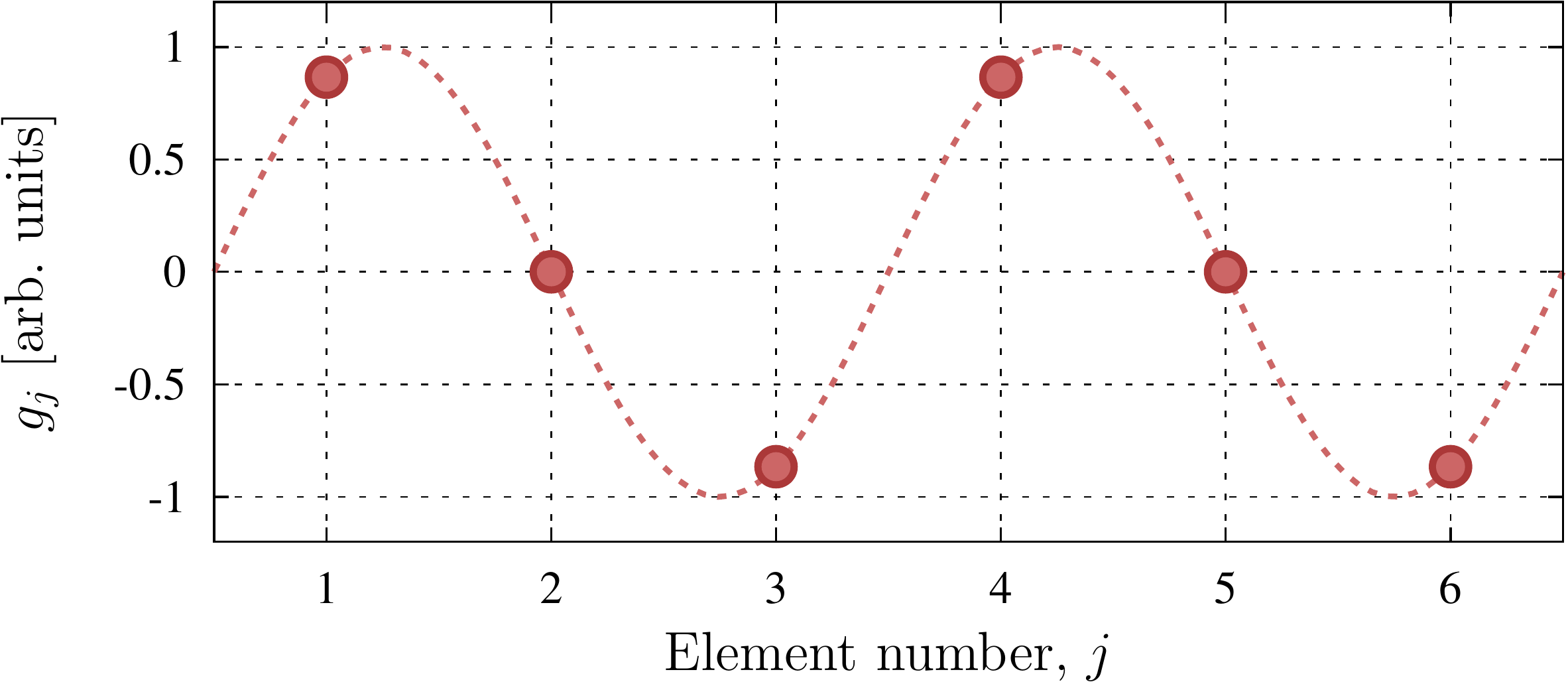}\\[2mm]%
 \includegraphics[width=\figurewidth]{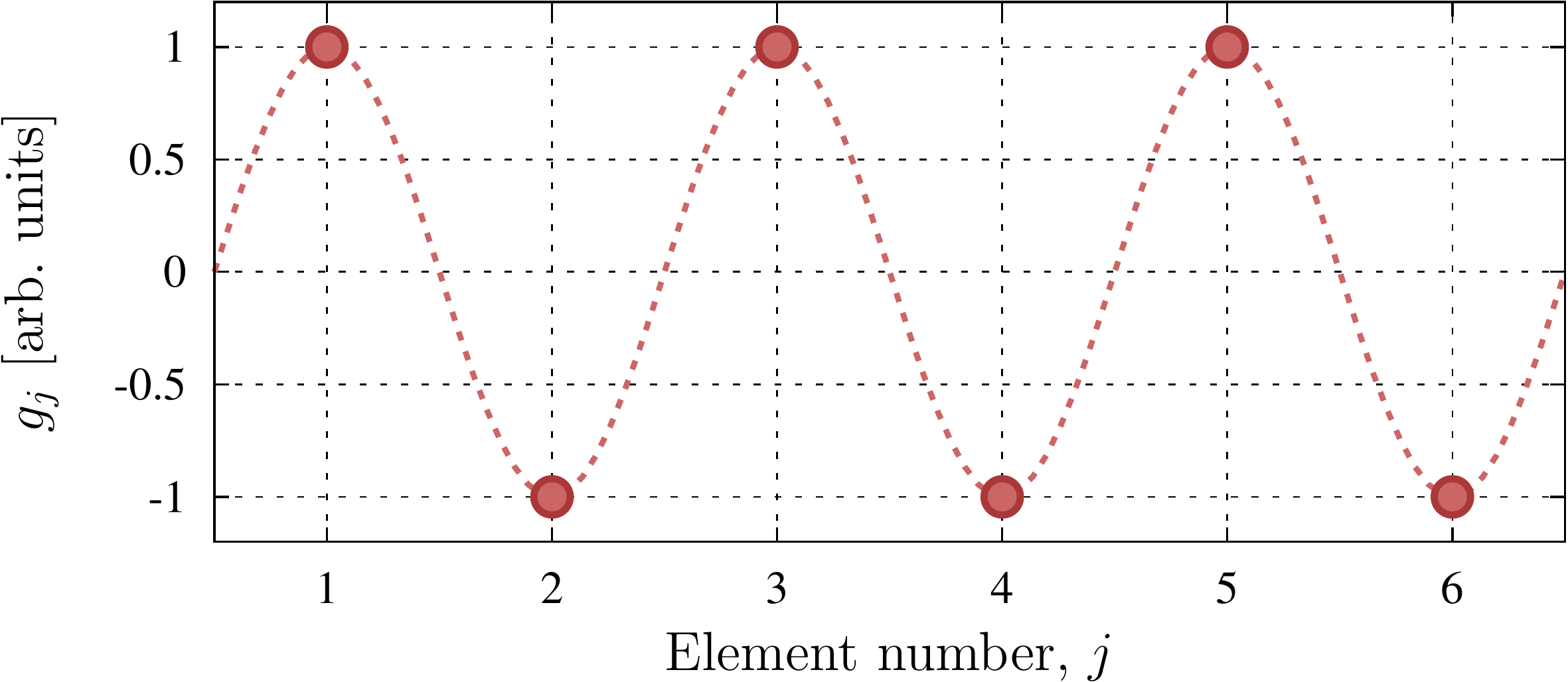}\\[2mm]%
 \includegraphics[width=\figurewidth]{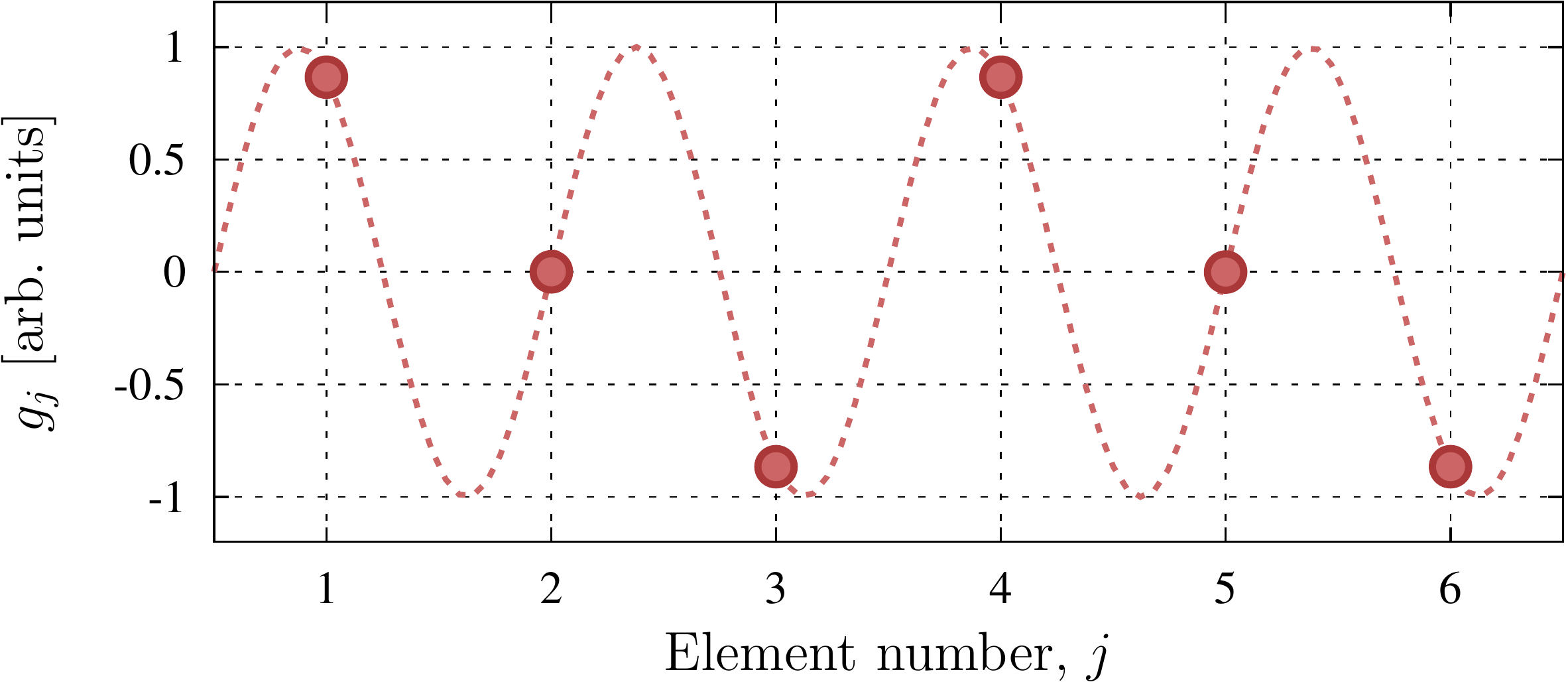}\\[2mm]%
 \includegraphics[width=\figurewidth]{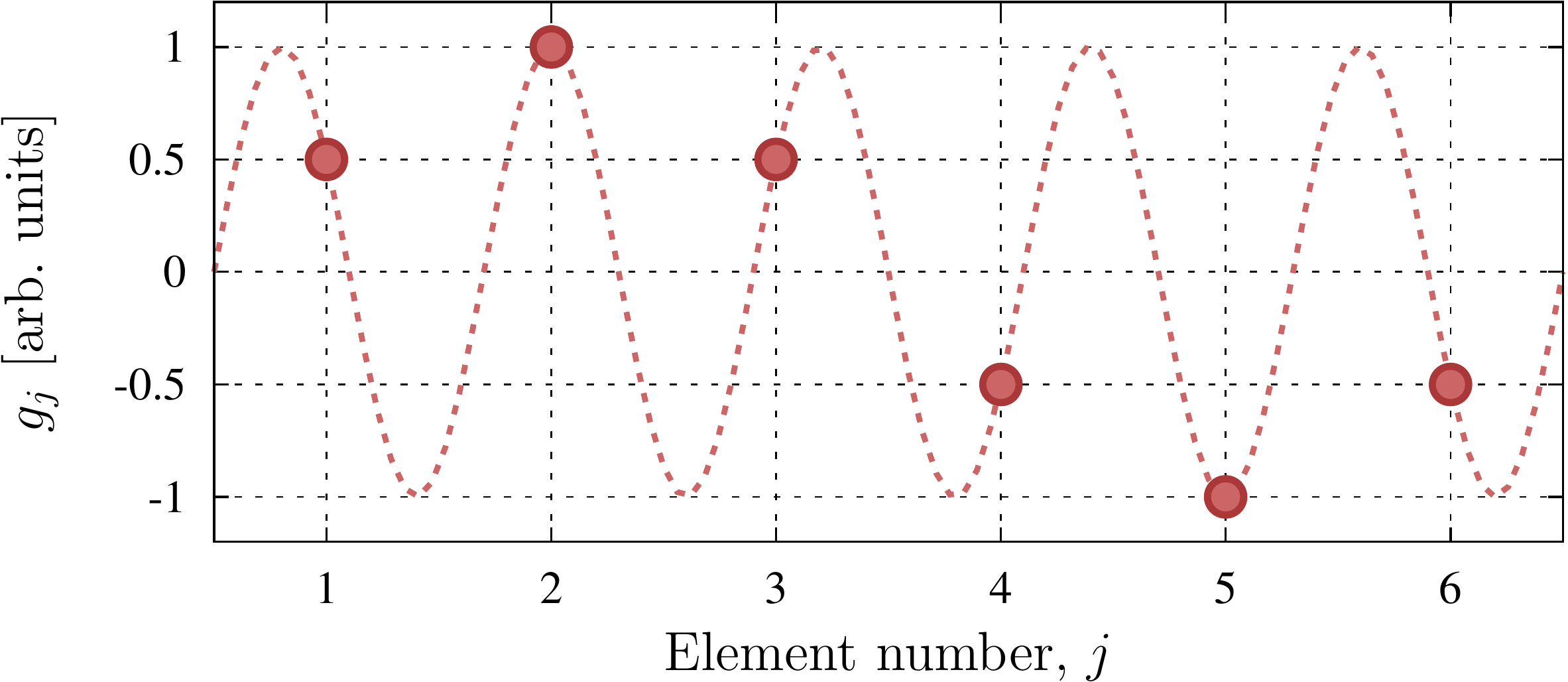}%
\caption{Individual coupling strengths for the case of $6$ elements ($l=1,\dots,5$, top to bottom); the dashed curves are drawn as guides to the eye.}%
 \label{fig:Profiles}%
\end{figure}%
\begin{equation}
\frac{\partial k}{\partial\delta x_j}=-\frac{\im{\beta+e^{-i\mu}\alpha}}{L+2d\frac{\partial\chi}{\partial\nu}}\,,
\end{equation}
with
\begin{equation}
\alpha=2ik\zeta\Bigl[e^{i\mu_1}(1+i\chi_1)\chi_2-e^{i\mu_2}\chi_1(1+i\chi_2)\Bigr]\,,
\end{equation}
and
\begin{equation}
\beta=2k\zeta\Bigl[\chi_1\chi_2-(1+i\chi_1)(1-i\chi_2)e^{i(\mu_1-\mu_2)}\Bigr]\,,
\end{equation}
which we can rewrite, by expressing $\chi_{1,2}$ and $\mu_{1,2}$ in terms of Chebyshev polynomials, as
\begin{align}
\alpha&=2ik\zeta^2\Biggl[\frac{(1+\zeta^2)U_{n_1-1}^2(a)U_{n_2-1}(a)}{(1-i\zeta)U_{n_1-1}(a)-e^{i\nu}U_{n_1-2}(a)}\nonumber\\
&\phantom{=2ik\zeta^2\Biggl[}\quad-\frac{(1+\zeta^2)U_{n_2-1}^2(a)U_{n_1-1}(a)}{(1-i\zeta)U_{n_2-1}(a)-e^{i\nu}U_{n_2-2}(a)}\Biggr]\,,
\end{align}
and
\begin{align}
\beta&=2k\zeta \biggl\{\zeta^2U_{n_1-1}(a)U_{n_2-1}(a)-\bigl[1+\zeta^2U_{n_1-1}^2(a)\bigr]\nonumber\\
&\phantom{=2k\zeta\biggl\{}\qquad\times\frac{(1-i\zeta)U_{n_2-1}(a)-e^{i\nu}U_{n_2-2}(a)}{(1-i\zeta)U_{n_1-1}(a)-e^{i\nu}U_{n_1-2}(a)}\biggr\}\,.
\end{align}
When $d=d_l$, one can show that these two expressions simplify considerably to yield
\begin{align}
\im{\beta+e^{-i\mu}\alpha}&\varpropto\sin\biggl(2l\pi\frac{j-\tfrac{1}{2}}{N}\biggr)\,,
\end{align}
This means that the individual membrane linear optomechanical couplings for the $l$-th transmissive point have a sinusoidal dependence with respect to their position in the array:
 \begin{align}
g_j^{(l)}&\varpropto\sin\biggl(2l\pi\frac{j-\tfrac{1}{2}}{N}\biggr)\,,
\end{align}
\begin{figure}[t]%
 \includegraphics[width=\figurewidth]{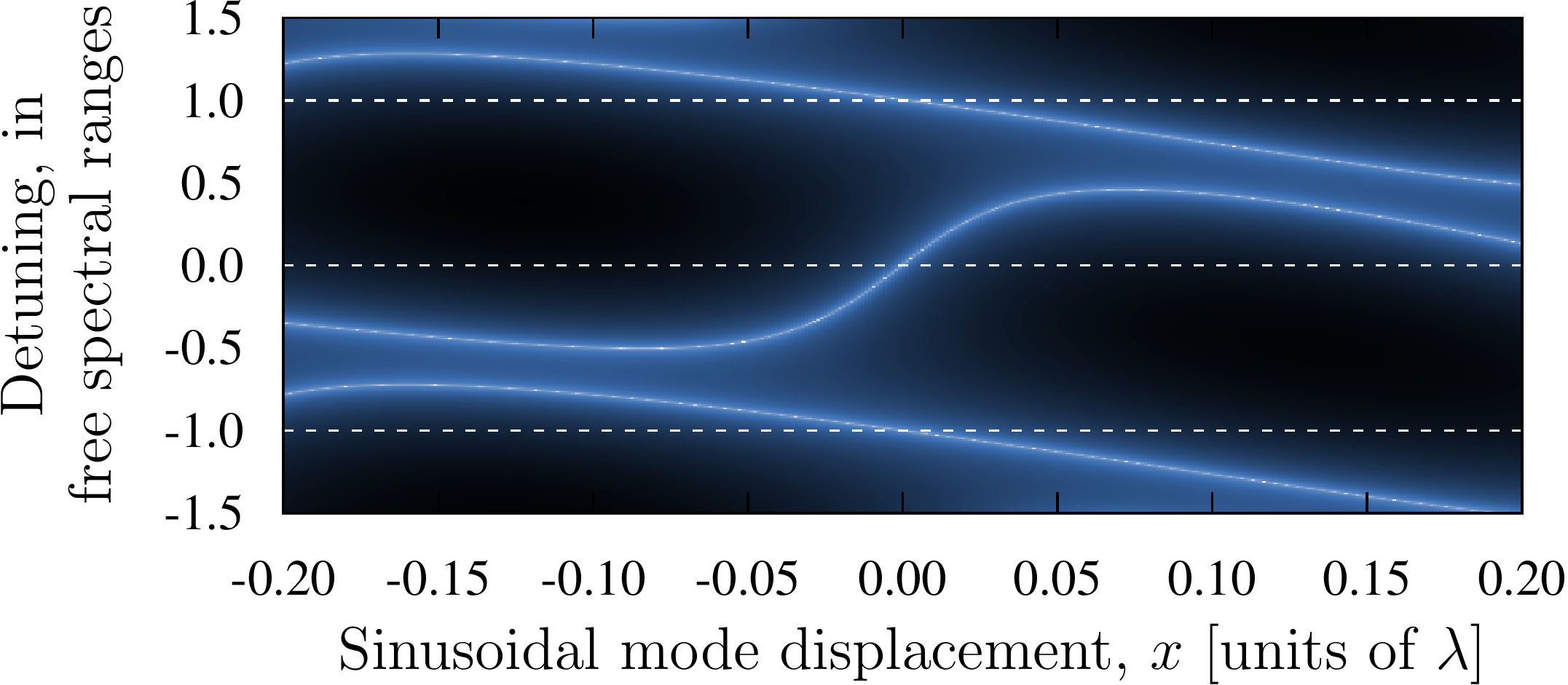}\\[1mm]%
 \includegraphics[width=\figurewidth]{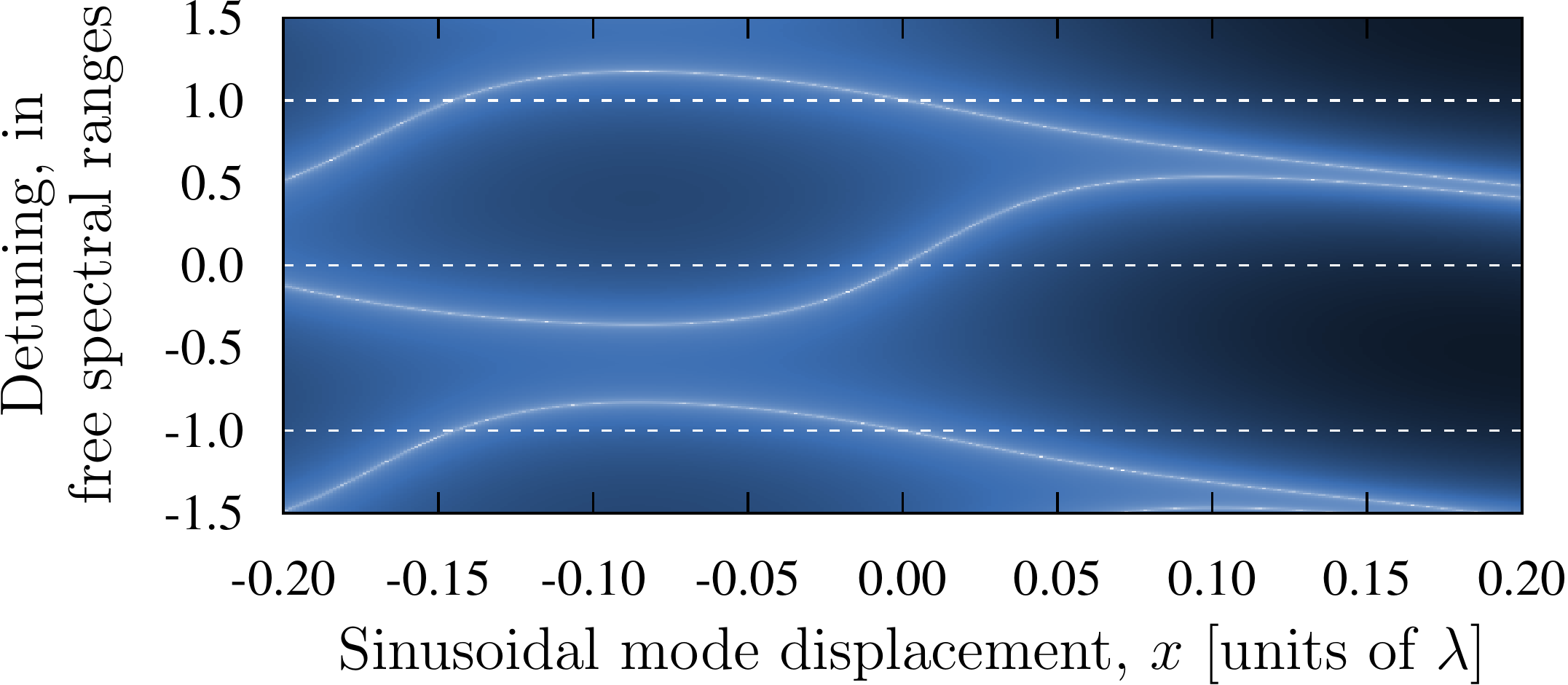}\\[1mm]%
 \includegraphics[width=\figurewidth]{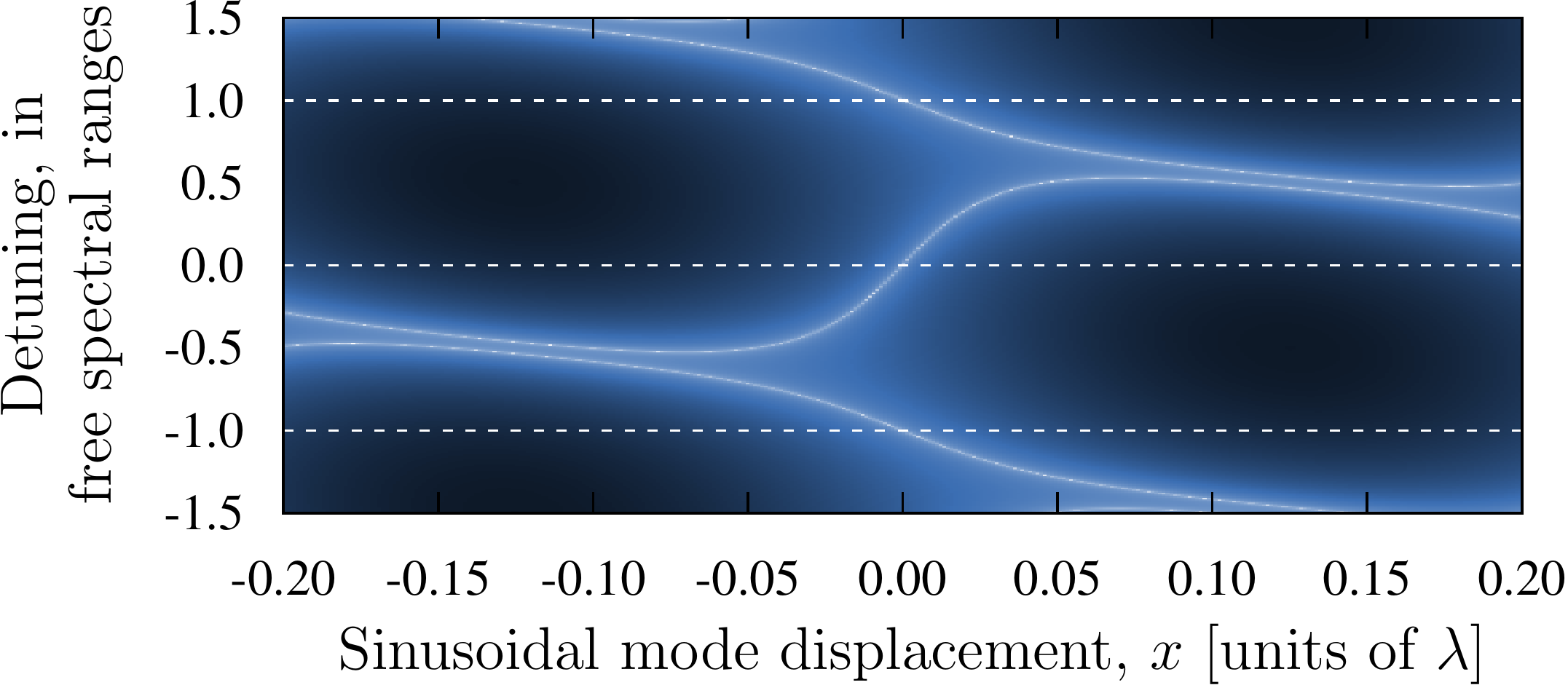}\\[1mm]%
 \includegraphics[width=\figurewidth]{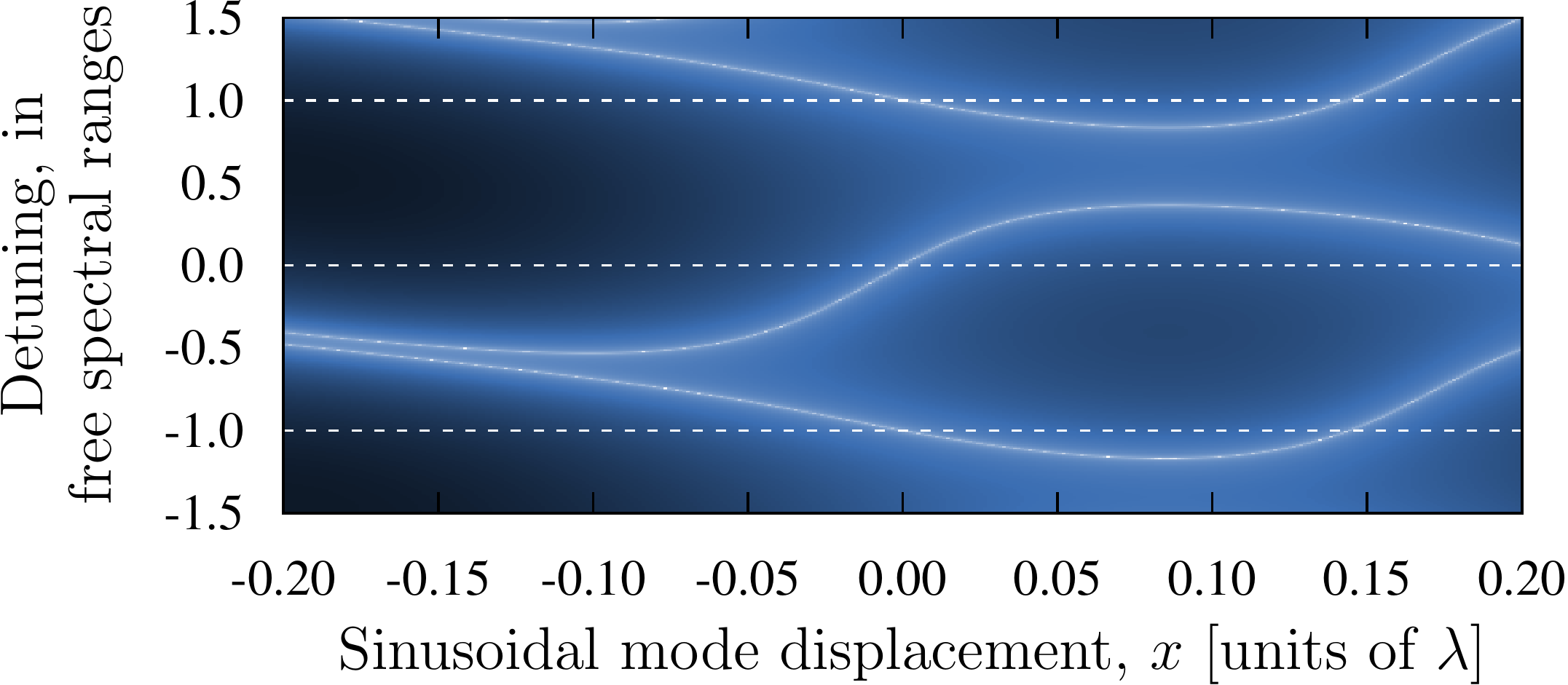}\\[1mm]%
 \includegraphics[width=\figurewidth]{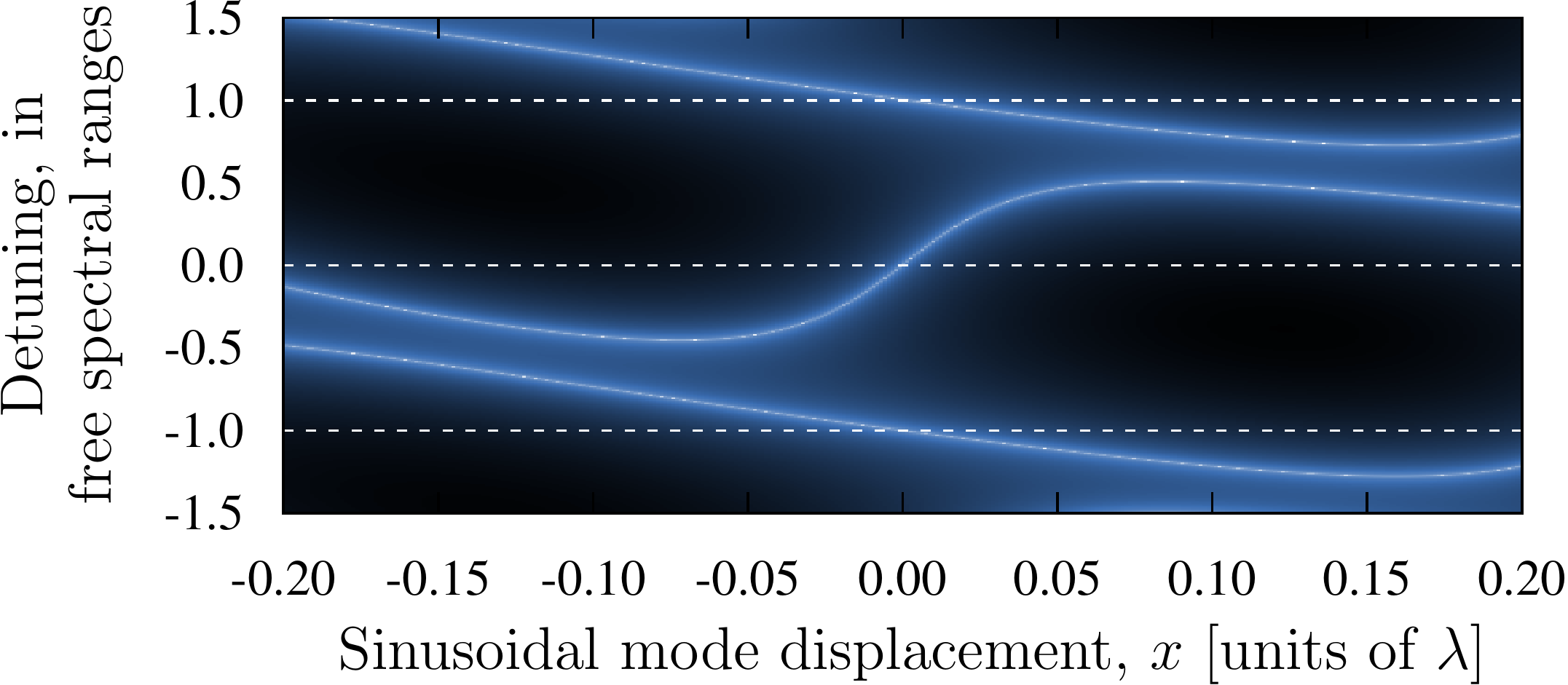}%
\caption{Transmission through a cavity with $N=6$ immobile elements close to the five transmission points; from top to bottom, the figures depict $l=1$ to $l=5$. This figure should be compared to \fref{fig:TransmissionCOM}. $x$ is normalized by a factor $\sqrt{N/2}$. (Parameters as in \fref{fig:TransmissionCOM}.)}%
 \label{fig:TransmissionSin}%
\end{figure}%
We illustrate the `profiles' of the $g_j^{(l)}$ for $N=6$ in \fref{fig:Profiles}. As we shall show below, the coupling of the collective motion of the membranes to the cavity field close to the $l$-th transmission point is governed by the constant
\begin{equation}
g_\mathrm{sin}^{(l)}\equiv\sqrt{\sum_{j=1}^N\bigl(g_j^{(l)}\bigr)^2}\,.
\end{equation}
Plots of the type of \fref{fig:TransmissionSin}, similarly to \fref{fig:TransmissionCOM} but shown here for the five sinusoidal modes that can be excited in an ensemble with $N=6$, can be used to numerically extract $g_\mathrm{sin}^{(l)}$ by measuring the gradient of the bright curves. Such data agree very well with the analytic results derived below.

\subsection{Collective motional-mode treatment: Heisenberg--Langevin formalism}
In the preceding section we derived coupling constants $g_j^{(l)}$ that relate the motion of the $j$\textsuperscript{th} element to the resonance frequency of the cavity at the $l$\textsuperscript{th} transmission point. A particular feature of multi-element arrays is that the cavity field couples to a \emph{collective} motion of the elements, with the $g_j^{(l)}$ playing the role of choosing the `profile' of the mode that is coupled to the cavity field, in the spirit of \fref{fig:Profiles}. To see this, let us describe the motion of the $j$\textsuperscript{th} mechanical element ($1\leq j\leq N$) through the annihilation operator $\hat{b}_j^{(l)}$, which obeys the Heisenberg--Langevin equation of motion~\cite{Giovannetti2001,Genes2008}
\begin{equation}
\tfrac{\rmd}{\rmd t}\hat{b}_j^{(l)}=-(i\omega_\mathrm{m}+\Gamma)\hat{b}_j^{(l)}+\hat{F}_j^{(l)}+\sqrt{2\Gamma}\hat{\xi}_j^{(l)}\,,
\end{equation}
where $\hat{\xi}_j^{(l)}$ is the relevant Langevin noise term whose properties we leave unspecified. For simplicity, we assume that all the oscillators have identical oscillation frequency $\omega_\mathrm{m}$, decay rate $\Gamma$, and temperature $T$, such that in thermal equilibrium they all have the same average occupation. $\hat{F}_j^{(l)}=g_j^{(l)}\hat{F}^{(l)}$ is a force term due to the action of the cavity, whose exact form is not relevant here. To describe the collective motion, we use the vector $(\tilde{g}_j^{(l)}\equiv g_j^{(l)}/g_\mathrm{sin}^{(l)})_j$, which is naturally normalized such that $\sum_{j=1}^N{\bigl(\tilde{g}_j^{(l)}\bigr)^2}=1$, and define:\ $\hat{b}^{(l)}\equiv\sum_{j=1}^N{\tilde{g}_j^{(l)}\hat{b}_j^{(l)}}$ and $\hat{\xi}^{(l)}\equiv\sum_{j=1}^N{\tilde{g}_j^{(l)}\hat{\xi}_j^{(l)}}$. Thus:
\begin{equation}
\tfrac{\rmd}{\rmd t}\hat{b}^{(l)}=-(i\omega_\mathrm{m}+\Gamma)\hat{b}^{(l)}+g_\mathrm{sin}^{(l)}\hat{F}^{(l)}+\sqrt{2\Gamma}\hat{\xi}^{(l)}\,.
\end{equation}
Under the assumption that the noise terms $\hat{\xi}_j^{(l)}$ are of a similar nature to one another and are independent (i.e., any cross-correlator between $\hat{\xi}_i^{(l)}$ and $\hat{\xi}_j^{(l)}$ is zero for $i\neq j$), then $\hat{\xi}^{(l)}$ obeys the same correlation functions as each \emph{individual} noise term, because of the normalization of $(\tilde{g}_j^{(l)})_j$, whereupon $\hat{b}^{(l)}$ behaves as a single collective oscillator with decay rate $\Gamma$. Let us remark at this point that our description in terms of this collective mode is one where we merely rotate to a different basis for this $N$-dimensional space, and therefore the correct normalization, necessary for the rotation to be a unitary operation, is indeed $\sum_{j=1}^N{\bigl(\tilde{g}_j^{(l)}\bigr)^2}=1$. Therefore, the dynamics of the cavity--mechanical system can be described entirely through an optomechanical Hamiltonian connecting a single cavity mode to a \emph{single collective mechanical mode} with coupling strength $g_\mathrm{sin}^{(l)}$, mechanical frequency $\omega_\mathrm{m}$, decay rate $\Gamma$, and noise operator $\hat{\xi}^{(l)}$.\\
In the next section we investigate some properties of this coupling strength $g_\mathrm{sin}^{(l)}$, and draw general conclusions regarding the optomechanical coupling of a transmissive ensemble.

\section{Transmissive optomechanics}
Thus far, we have derived an expression for the coupling strength of a cavity field to a periodic array of elements whose compound reflectivity is zero at the cavity frequency. In this section we shall derive analytical expressions for the resulting collective optomechanical coupling for the two outermost transmission points $l=1,N-1$. We shall then proceed to discuss one further important consequence of this collective coupling:\ a linewidth-narrowing effect where the effective linewidth of the cavity resonance decreases over its `bare' value as the number of elements, or their reflectivity, is increased. We will then proceed with an investigation of these effects for the other transmission points. 

\subsection{Enhanced optomechanical coupling}
As discussed above, the coupling of the collective motion of the membranes to the cavity field close to the $l$-th transmission point is governed by the constant $g_\mathrm{sin}^{(l)}$. Setting $l=1$ (essentially the same results will be obtained for $l=N-1$) one can show that, when $\zeta<0$,
\begin{multline}
g_j^{(1)}=-2\omega_\mathrm{c}x_0\frac{\zeta\csc\bigl(\frac{\pi}{N}\bigr)\Bigl[\sqrt{\sin^2\bigl(\frac{\pi}{N}\bigr)+\zeta^2}-\zeta\Bigr]}{L-2Nd\zeta\csc^2\bigl(\frac{\pi}{N}\bigr)\sqrt{\sin^2\bigl(\frac{\pi}{N}\bigr)+\zeta^2}}\\
\times\sin\biggl(2\pi\frac{j-\tfrac{1}{2}}{N}\biggr)\,.
\end{multline}
which yields, for $N=2$,
\begin{equation}
\label{eq:gsinN2}
g_\mathrm{sin}^{(1)}=-g\frac{\sqrt{2}\,\zeta\bigl(\sqrt{1+\zeta^2}-\zeta\bigr)}{1-4\tfrac{d}{L}\zeta\sqrt{1+\zeta^2}}\,,
\end{equation}
and for $N>2$
\begin{equation}
\label{eq:gsin}
g_\mathrm{sin}^{(1)}=-g\sqrt{\frac{N}{2}}\frac{\zeta\csc\bigl(\frac{\pi}{N}\bigr)\Bigl[\sqrt{\sin^2\bigl(\frac{\pi}{N}\bigr)+\zeta^2}-\zeta\Bigr]}{1-2N\tfrac{d}{L}\zeta\csc^2\bigl(\frac{\pi}{N}\bigr)\sqrt{\sin^2\bigl(\frac{\pi}{N}\bigr)+\zeta^2}}\,.
\end{equation}
These two expressions do not agree upon setting $N=2$ in the latter; this anomaly is due to the relation
\begin{equation}
\label{eq:NormalisationAnomaly}
\sqrt{\sum_{j=1}^N\sin^2\biggl(2\pi\frac{j-\tfrac{1}{2}}{N}\biggr)}=\begin{cases}
\sqrt{2} & \text{for }N=2\\
\sqrt{\frac{N}{2}} & \text{for}\ N>2
\end{cases}\,.
\end{equation}
The first thing we note is the fact that $g_\mathrm{sin}$ is no longer bounded above by $g$. Indeed, for large $N$ and $\abs{\zeta}$ but small $d/L$, the expression (\ref{eq:gsin}) for $g_\mathrm{sin}^{(1)}$ simplifies considerably to yield
\begin{equation}
\label{eq:SinusoidalCouplingSimplified}
g_\mathrm{sin}^{(1)}=g\frac{\tfrac{\sqrt{2}}{\pi}\,\zeta^2N^{3/2}}{1+\tfrac{2}{\pi^2}\tfrac{d}{L}\zeta^2N^3}\approx\frac{\sqrt{2}}{\pi}\,g\,\zeta^2N^{3/2}\,.
\end{equation}
We shall provide numerical examples later to show that $g_\mathrm{sin}^{(1)}$ can be orders of magnitude larger than $g$. To explore the scaling of $g_\mathrm{sin}^{(1)}/g$ with $N$ we begin by considering a very long cavity ($L\ggg d$) and approximate the denominator of \eref{eq:gsin} by $1$, obtaining
\begin{equation}
g_\mathrm{sin}^{(1)}=\sqrt{N/2}g\abs{\zeta}\bigl[\sqrt{1+(N\zeta/\pi)^2}-(N\zeta/\pi)\bigr]\,,
\end{equation}
for large $N$. One can now distinguish between two cases: (i)~$(N\abs{\zeta}/\pi)\ll1$:\ $g_\mathrm{sin}^{(1)}/g=\abs{\zeta}\sqrt{N/2}$, yielding the $\sqrt{N}$-scaling observed in, e.g., atom-cavity optomechanics experiments~\cite{Kruse2003,Murch2008,Brennecke2008,Schleier-Smith2011} that involve large ensembles of very low reflectivity scatterers. The coupling strength is also multiplied by a factor of $\abs{\zeta}$, which amplifies the interaction when $\abs{\zeta}>1$; both these features represent a markedly different behavior from the reflective regime. (ii)~$(N\abs{\zeta}/\pi)\gg1$, which gives the $N^{3/2}$- and $\zeta^2$-scaling shown in \eref{eq:SinusoidalCouplingSimplified}. This scaling with $N^{3/2}$ is a consequence of the modification of the field mode profile inside the cavity. Indeed, as $N$ increases the fraction of the energy density per photon inside the array increases strongly; an increased optomechanical coupling strength is consistent with this increase of energy density~\cite{Cheung2011}. The former case, on the other hand, corresponds to an essentially unperturbed cavity field mode, where there is no such concentration of energy density, and a weaker scaling with $N$ is therefore observed.\\
Similarly, in the behavior of $g_\mathrm{sin}^{(1)}/g$ as a function $\zeta$ for $N=2$ and in the same `long cavity' limit, one can distinguish between two regimes [cf.\ \eref{eq:gsinN2}, setting the denominator to $1$]. For $\abs{\zeta}\ll1$, $g_\mathrm{sin}^{(1)}/g$ grows linearly with $\abs{\zeta}$ as a consequence of the increased reflectivity of each element. For $\abs{\zeta}\gg1$, however, the cavity field mode is modified substantially and is strongly concentrated in the region between the two elements. This concentration grows quickly as a function of $\abs{\zeta}$ and gives rise to a quadratic scaling of $g_\mathrm{sin}^{(1)}/g$ with $\abs{\zeta}$.\\
\begin{figure}[t]%
 \includegraphics[width=\figurewidth]{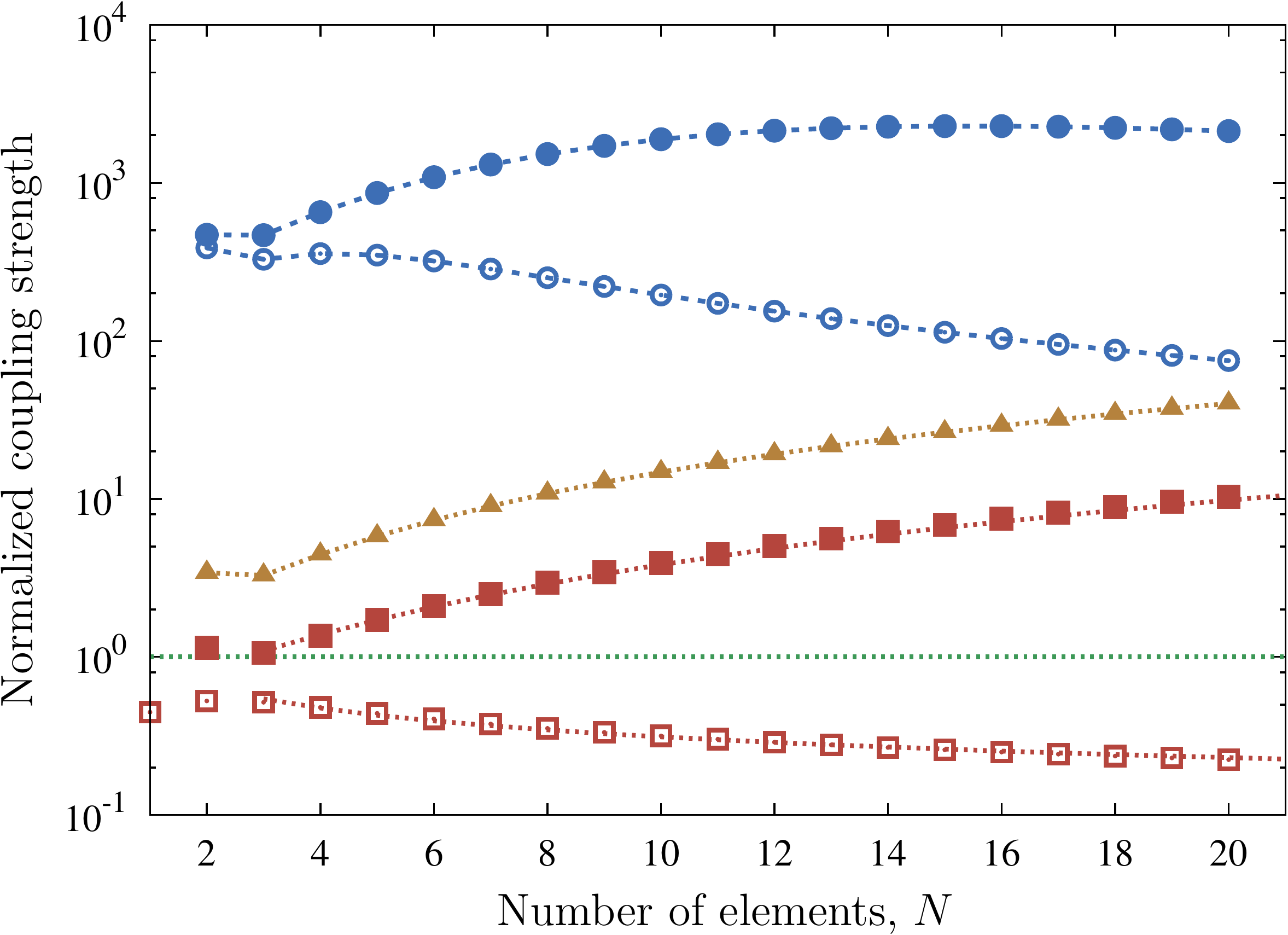}%
\caption{Collective coupling strength $g_\mathrm{sin}^{(1)}$ (blue circles, closed red squares, orange triangles) for a cavity of length $L\approx6.3\times10^4\lambda$ and several choices of scatterer reflectivity, compared to the center-of-mass coupling strength $g_\mathrm{com}$ (open red squares) and $g$ (green line). For the red, orange, and blue data points we choose a per-element intensity reflectivity of $20$\%, $50$\%, and $99.4$\%, respectively. For the center-of-mass data we illustrate the $N^{-1/2}$ scaling law that applies for large $N$, whereas for the first sinusoidal mode we draw a curve through the data points as a guide to the eye. Throughout this plot we take $d=d_1$, except for the data denoted by the open blue circles, for which $d=d_1+20\lambda$.}%
 \label{fig:Coupling}%
\end{figure}%
The denominator in the full form of \eref{eq:SinusoidalCouplingSimplified} can be interpreted as an effective renormalization of the cavity length from $L$ to
\begin{equation}
L_\mathrm{eff}^{(1)}\equiv L-2Nd\zeta\csc^2\biggl(\frac{\pi}{N}\biggr)\sqrt{\sin^2\biggl(\frac{\pi}{N}\biggr)+\zeta^2}\,;
\end{equation}
in the same regime as for \eref{eq:SinusoidalCouplingSimplified}, $L_\mathrm{eff}^{(1)}\approx L+\tfrac{2}{\pi^2}d\zeta^2N^3$. We shall discuss the regime in which the effective length is significantly larger than $L$, where an interesting linewidth-narrowing effect occurs, in greater detail in \sref{sec:Trans:LN} below. We have already seen that for $N$, $\abs{\zeta}$, and $d/L$ small enough that $L_\mathrm{eff}^{(1)}\approx L$ the coupling strength scales as $\zeta^2N^{3/2}$. On the other hand, when the parameters are such that $L_\mathrm{eff}^{(1)}\gg L$, $g_\mathrm{sin}^{(1)}$ does not depend on $\zeta$ and decreases as $N^{-3/2}$. We say that, in transitioning between the two scaling laws, $g_\mathrm{sin}$ \emph{saturates} (i.e., reaches a maximum value at some finite value for $N$) before it starts decreasing.\\
Optimizing $g_\mathrm{sin}^{(1)}$ over $N$ for arbitrary $L/d$, in this manner, we obtain $g_\mathrm{opt}^{(1)}=\tfrac{1}{2}g\sqrt{L/d}\abs{\zeta}$. This expression is valid for $\abs{\zeta}$ that is not too large, since the optimal number of elements must be $>2$. This favorable scaling with both $N$ and $\abs{\zeta}$ is a significant improvement over the state of the art. Close inspection reveals that $g_\mathrm{opt}^{(1)}$ is proportional to $1\big/\sqrt{Ld}$ and therefore can be improved either by making the main cavity smaller (i.e., decreasing $L$) or, independently, by positioning the elements closer together (decreasing $d$ to a smaller value whilst maintaining the condition of zero reflectivity).\\
\begin{figure}[t]%
 \includegraphics[width=\figurewidth]{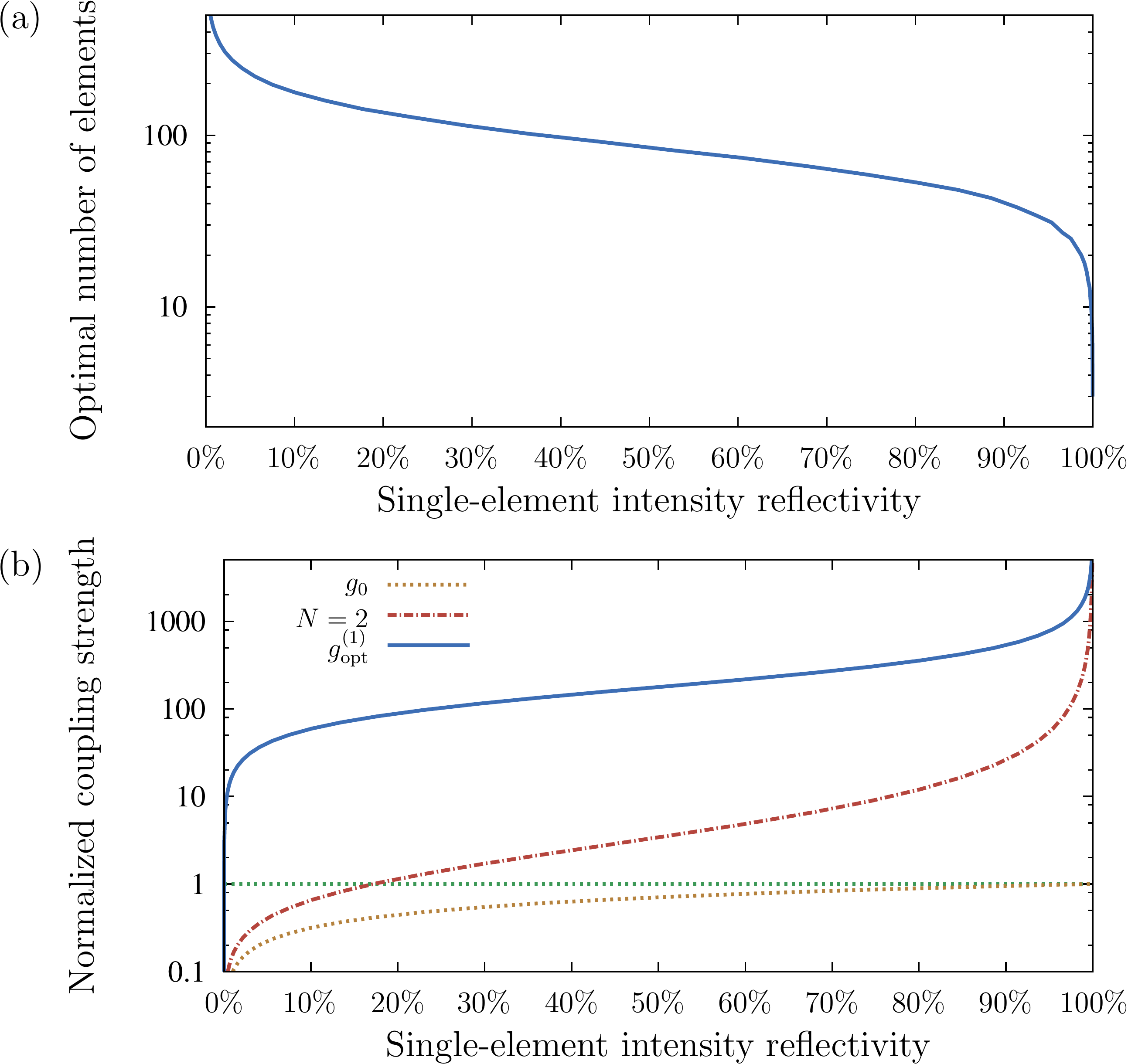}%
\caption{(a)~Optimal number of elements $N_\mathrm{opt}^{(1)}$, as a function of the single-element reflectivity. For reflectivities close to $100$\%, this number decreases rather quickly to $\lesssim10$. (b)~The coupling of the first sinusoidal mode, normalized to $g$, optimized as $N=N_\mathrm{opt}^{(1)}$ (solid blue curve), compared to the case for $N=2$ (dashed-dotted red) and $g_\mathrm{com}$ (dotted orange). For these plots we used $L\approx6.3\times10^4\lambda$ and $d=d_1$.}%
 \label{fig:OptElts}%
\end{figure}%
Alternatively, one may optimize the parameters such that the second term in the denominator of \eref{eq:SinusoidalCouplingSimplified} dominates, i.e., where $L_\mathrm{eff}^{(1)}\gg L$. The coupling strength then takes the approximate form
\begin{equation}
\frac{\pi}{\sqrt{2}}\,g\frac{L}{d}\,N^{-3/2}\xrightarrow[\ N=2\ ]{}\tfrac{\pi}{2}g_\mathrm{sm}\,,
\end{equation}
where we have taken the optimal ($N=2$) case and defined $g_\mathrm{sm}\equiv\omega_\mathrm{c}x_\mathrm{zpt}/d$ as the optomechanical coupling strength for a (small) cavity of length $d$ with a single moving mirror. In this regime, therefore, the system acts as a small cavity of length $d$ and is sensitive to relative motion between the two elements but not to the length of the main cavity.
\Fref{fig:Coupling} illustrates, primarily, the enhancement of optomechanical coupling strength that can be obtained by operating in the transmissive regime as compared to coupling to the center-of-mass motion. Several important observations can be made from this figure. The largest value for $g_\mathrm{sin}^{(1)}$ increases with $\abs{\zeta}$; in the figure we show data points for $\zeta=-0.5$ ($20$\% intensity reflectivity; red data marked with squares), $\zeta=-1.0$ ($50$\%; orange, triangles) and $\zeta=-12.9$ ($99.4$\%; blue, circles), the first two of which represents a typical reflectivity for SiN membranes used for optomechanical experiments~\cite{Thompson2008}, and the last membranes with increased reflectivity due to the use of sub-wavelength patterning~\cite{Kemiktarak2012,Kemiktarak2012b}. Secondly, the value of $N$ for which the coupling strength is optimized is highly dependent on the value of $d$; the reflectivity of the array depends on $d\mod(\lambda/2)$, so that one is free to increase the element spacing by integer multiples of half a wavelength without affecting its transmission properties (in the limit of a wavelength-independent reflectivity). However, the coupling strength is sensitive to this increase: The larger $d/L$ is, the earlier the \emph{saturation point} is reached, beyond which increasing $N$ lowers the coupling strength. The figure also illustrates two scaling laws that we derived above. The center-of-mass coupling decreases, for $N\gtrsim3$, as $N^{-1/2}$, whereas the sinusoidal coupling strength $g_\mathrm{sin}^{(1)}$ scales approximately as $N^{3/2}$ for $N$ large enough, but also small enough to avoid the effects of saturation. These two scaling laws are illustrated by the red dotted curves, drawn as guides to the eye. By contrast, the other curves and all the data points are generated using the full analytical formulae, which are in excellent agreement with numeric calculations.\\
In \fref{fig:OptElts} we study (a)~the optimal number of elements required for $l=1$, $N_\mathrm{opt}^{(1)}$, and (b)~the resulting coupling, as a function of the polarizability of each element in the array. For weakly reflective elements, as illustrated in \fref{fig:OptElts}(a), the coupling only saturates for very large values of $N$, whereas as the reflectivity of the elements increases the optimal number of elements decreases, at first steadily and then quite sharply, until it reaches a point where the coupling strength decreases for $N>2$. This curve is sensitive to the ratio $d/L$; smaller values of this ratio result in larger values for $N_\mathrm{opt}^{(1)}$. In panel~(b), we illustrate the optimized sinusoidal coupling strength (solid blue curve), as well as the coupling strength for $N=2$ (dashed-dotted red) and the center-of-mass coupling (dotted orange). Two immediate observations can be made that are quite general. First, $g_\mathrm{sin}^{(1)}$ may exceed $g_\mathrm{com}$ by several orders of magnitude. Second, in the case of weakly-reflective elements, it is necessary to use rather large values for $N$ to achieve this orders-of-magnitude improvement in coupling strength.

\begin{figure}[t]%
 \includegraphics[width=\figurewidth]{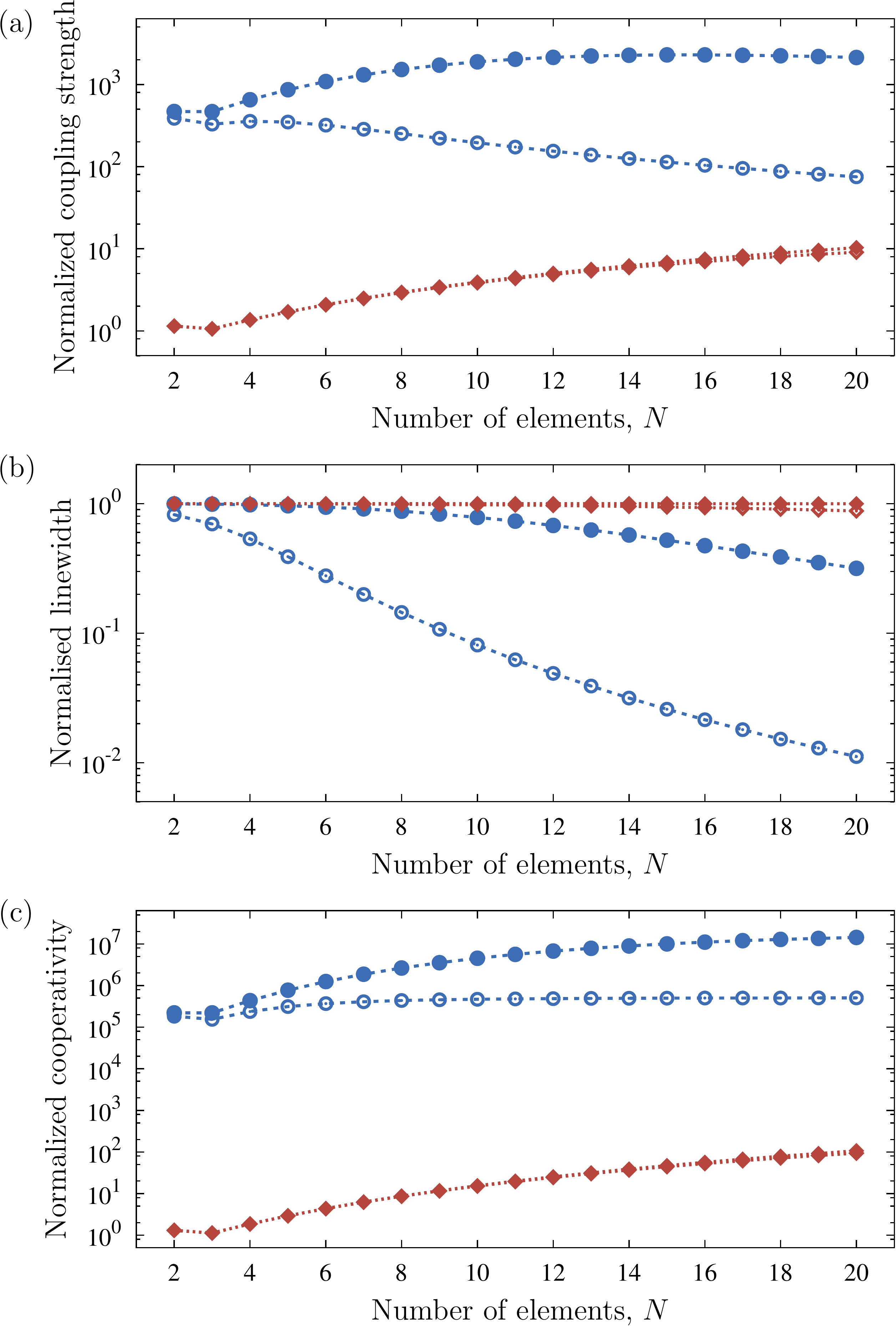}%
\caption{(a)~Coupling strength $g_{\mathrm{sin}}^{(1)}$, normalized to $g$, as a function of the number of elements. Four sets of data are shown. The red diamonds (blue circles) correspond to a single-element reflectivity of $20$\% ($99.4$\%). Closed (open) symbols represent an inter-element spacing of $d=d_1$ ($d=d_1+20\lambda$). We note that the inter-element separation has a much stronger effect for highly-reflective elements. ($L\approx6.3\times10^4\lambda$, bare-cavity finesse $\approx\,3\times10^4$.) (b)~A similar plot showing the effective cavity linewidth, normalized to the bare-cavity linewidth, obtained in each of the cases displayed in the upper plot. (c)~Putting these two together we can calculate the cooperativity, normalized to the single-element cooperativity, and demonstrate an enhancement by several orders of magnitude for the chosen parameters.}%
 \label{fig:LineCoop}%
\end{figure}%
\subsection{Linewidth narrowing}\label{sec:Trans:LN}
The saturation phenomenon described above reveals another interesting effect as the number of elements is increased beyond $N_\mathrm{opt}^{(1)}$: The presence of the array inside the cavity acts to narrow the cavity linewidth. The physical basis behind this is rather transparent and relies on two observations. First, the fact that the array is transparent at the cavity resonance frequency means that the finesse of the cavity -- which is related to the number of round-trips a photon makes inside the cavity on average -- \emph{is unchanged} by its presence. The second observation is that, as we have already noted, \eref{eq:SinusoidalCouplingSimplified} reveals that the cavity is effectively lengthened to a length $L_\mathrm{eff}^{(1)}$. Since the bare-cavity linewidth is $\kappa_\mathrm{c}\varpropto1/(\mathcal{F}L)$, it follows that the linewidth of the cavity is reduced to $\kappa_\mathrm{eff}^{(1)}\varpropto1/(\mathcal{F}L_\mathrm{eff}^{(1)})$. By choosing the right parameters, one can optimize for this linewidth-narrowing effect, \fref{fig:LineCoop}(b), to narrow the optical resonance substantially. As several mechanisms in optomechanics, e.g., cooling in the linearized regime, are improved in the so-called `resolved sideband' regime, where $\omega_\mathrm{m}\gg\kappa_\mathrm{c}$, the collective mechanism we describe can result in the condition $\kappa_\mathrm{c}>\omega_\mathrm{m}\gg\kappa_\mathrm{eff}^{(1)}$ being satisfied, thereby improving the performance of these mechanisms in the system.
\par
One other figure of merit that is relevant to several mechanisms is the cooperativity $C=g^2/(\kappa\Gamma)$, i.e., the ratio of the square of the optomechanical coupling strength to the product of the optical and mechanical decay rates, $\kappa$ and $\Gamma$, respectively. For a multi-element system composed of independent oscillators and operating in the transmissive regime, $\Gamma$ is independent of $N$, but $g_\mathrm{sin}^{(1)}\sim N^{3/2}$ and $\kappa_\mathrm{c}\sim N^3$ in the appropriate regime; this is illustrated in \fref{fig:LineCoop}(a) and~(b). The result is a competition between these two factors, yielding a constant cooperativity as $N$ is increased. In \fref{fig:LineCoop}(c) we plot the \emph{normalized} cooperativity, i.e., the cooperativity for the $N$-element ensemble divided by that for a single element; the enhancement obtained with the parameters used is of almost $10^7$. Even when absorption is included, an enhancement of several orders of magnitude is still possible~\cite{Xuereb2012c}.

\subsection{Other transmission points}
\begin{figure}[t]%
 \includegraphics[width=\figurewidth]{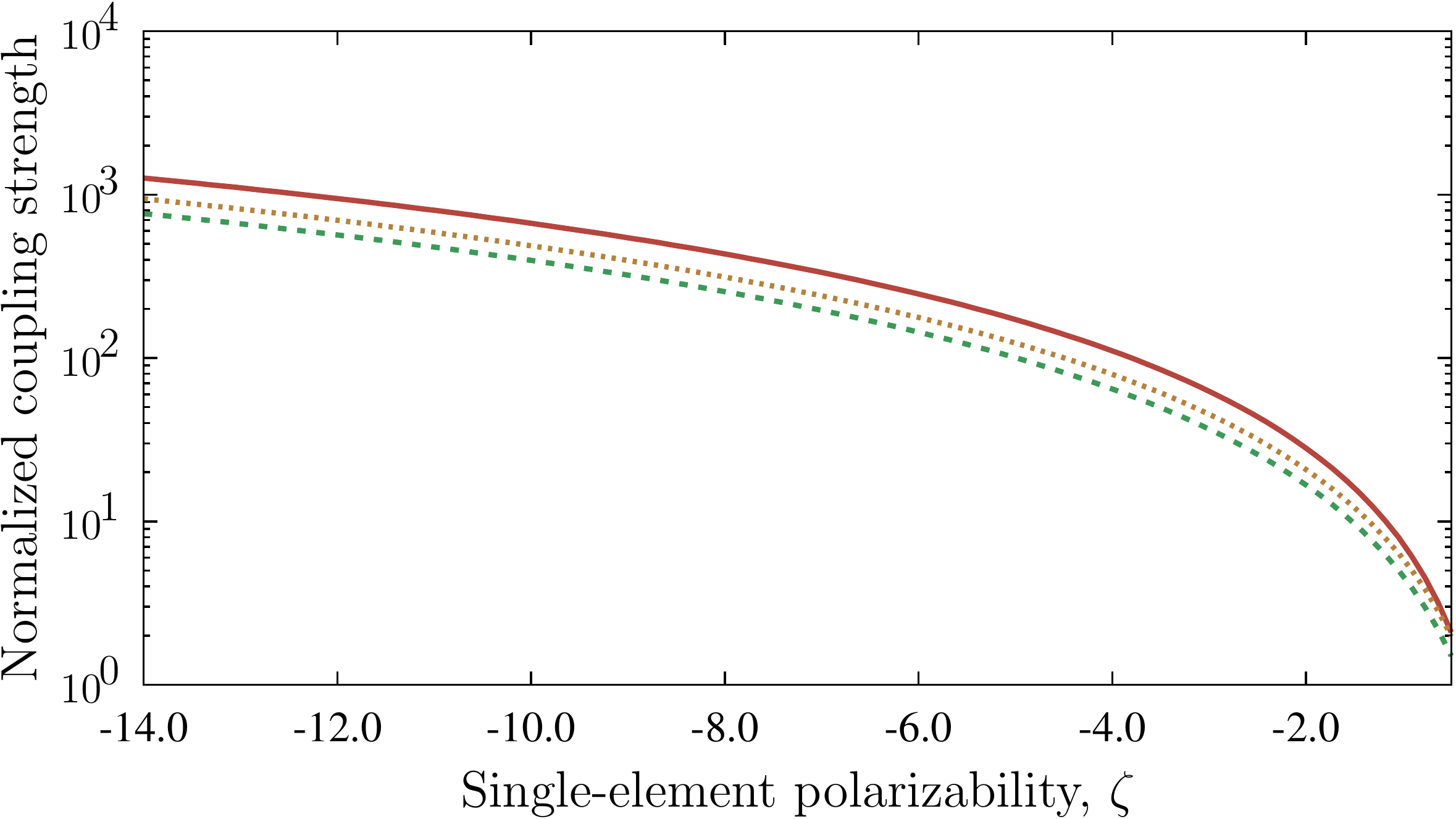}%
\caption{Coupling strength $g_{\mathrm{sin}}^{(l)}$, normalized to $g$, as a function of $\zeta$, for $N=6$. Three curves are shown:\ $d=d_1$ (solid red), $d=d_2$ (dashed green), and $d=d_3$ (dotted orange). ($L\approx6.3\times10^4\lambda$, bare-cavity finesse $\approx\,3\times10^4$.)}%
 \label{fig:ScalingCouplingZeta}%
\end{figure}%
\begin{figure}[t]%
 \includegraphics[width=\figurewidth]{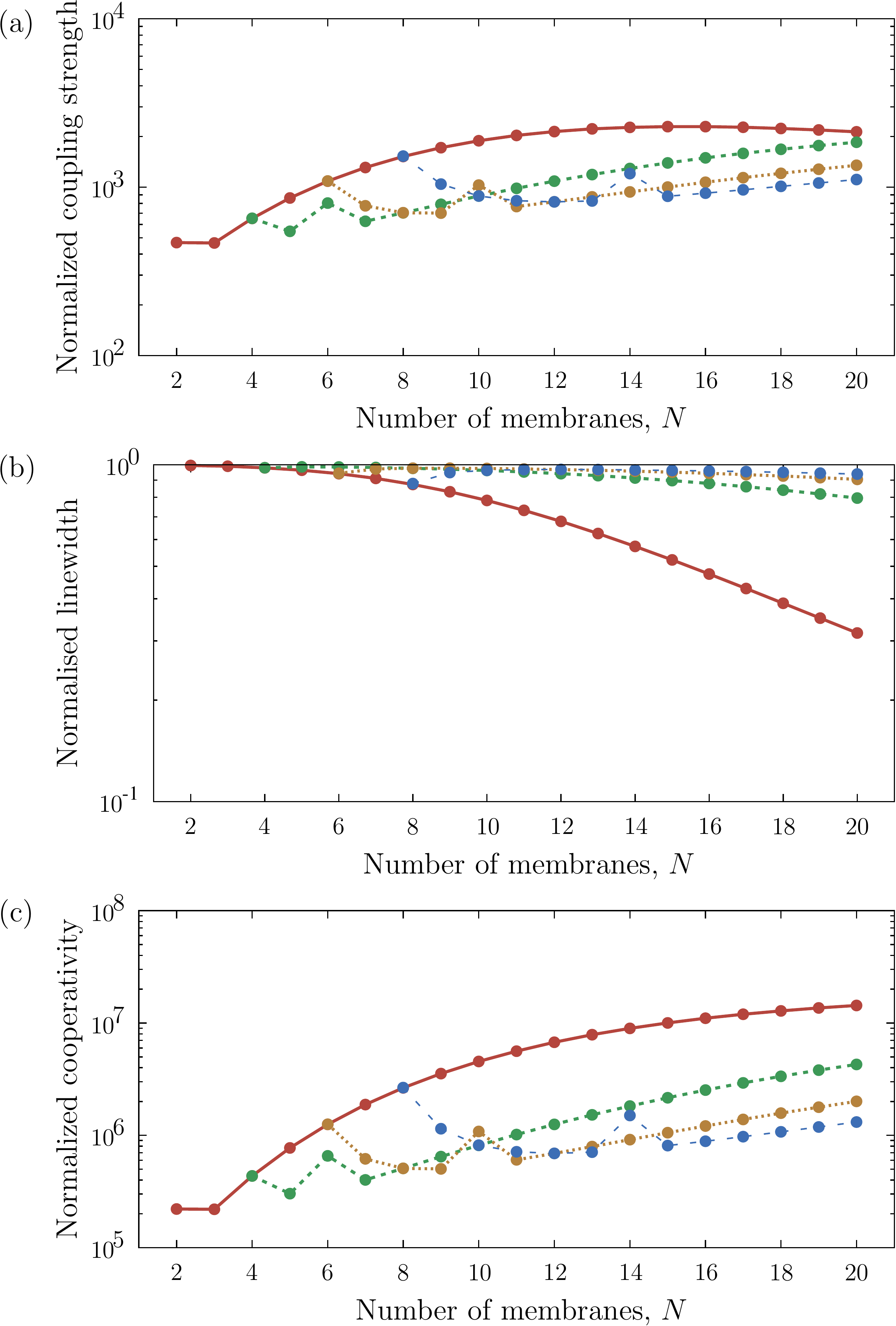}%
\caption{(a)~Coupling strength $g_{\mathrm{sin}}^{(l)}$, normalized to $g$, (b)~effective linewidth, normalized to the bare-cavity linewidth, and (c)~cooperativity, normalized to the single-element cooperativity, all plotted as functions of $N$, for $\zeta=-12.9$. Four curves are shown in each panel:\ $d=d_1$ (solid red), $d=d_3$ (dashed green), $d=d_5$ (dotted orange) and $d=d_7$ (long-dashed blue). The red curve in each part is to be compared to the corresponding solid blue data points in \fref{fig:LineCoop}. See also \fref{fig:ScalingD20}. ($L\approx6.3\times10^4\lambda$, bare-cavity finesse $\approx\,3\times10^4$.)}%
 \label{fig:ScalingD0}%
\end{figure}%
\begin{figure}[t]%
 \includegraphics[width=\figurewidth]{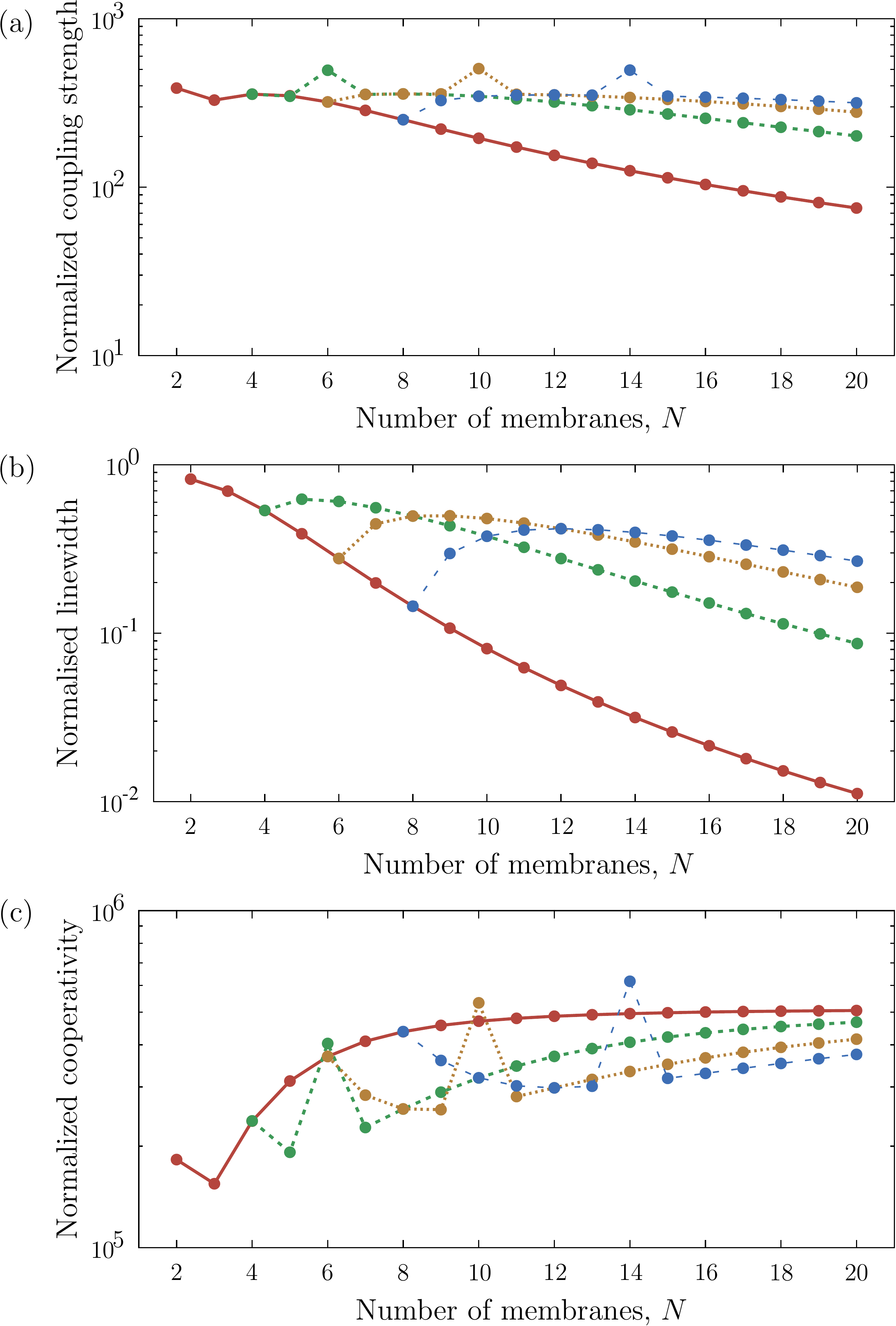}%
\caption{Similar to \fref{fig:ScalingD0}, but the four curves shown in each panel are for:\ $d=20\lambda+d_1$ (solid red), $d=20\lambda+d_3$ (dashed green), $d=20\lambda+d_5$ (dotted orange) and $d=20\lambda+d_7$ (long-dashed blue). The red curve in each part is to be compared to the corresponding open blue data points in \fref{fig:LineCoop}. (Other parameters as in \fref{fig:ScalingD0}.)}%
 \label{fig:ScalingD20}%
\end{figure}%
Let us now turn to the other transmissive points, indexed by $1\leq l\leq N-1$. Analytical expressions for the coupling strength at transmission points other than $l=1,N-1$ could not be derived easily. Within the numerical precision of our simulations, however, we found excellent agreement with the expressions ($N>2$)
\begin{equation}
g_\mathrm{sin}^{(l)}=-\mathcal{N}_{N,l}g\frac{\zeta\csc\bigl(\frac{l\pi}{N}\bigr)\Bigl[\sqrt{\sin^2\bigl(\frac{l\pi}{N}\bigr)+\zeta^2}-\zeta\Bigr]}{1-2N\tfrac{d}{L}\zeta\csc^2\bigl(\frac{l\pi}{N}\bigr)\sqrt{\sin^2\bigl(\frac{l\pi}{N}\bigr)+\zeta^2}}\,,
\end{equation}
where
\begin{equation}
\mathcal{N}_{N,l}=\begin{cases}
\sqrt{\frac{N}{2}} & \text{for }N\neq2l\\
\sqrt{N} & \text{for }N=2l\\
\end{cases}\,,
\end{equation}
thereby yielding an effective length
\begin{equation}
L_\mathrm{eff}^{(l)}\equiv L-2Nd\zeta\csc^2\biggl(\frac{l\pi}{N}\biggr)\sqrt{\sin^2\biggl(\frac{l\pi}{N}\biggr)+\zeta^2}\,.
\end{equation}
The effective linewidth of the system can then be written as
\begin{align}
\kappa_\mathrm{eff}^{(l)}&=\frac{c}{2L_\mathrm{eff}^{(l)}}\frac{1}{\abs{Z}\sqrt{Z^2+1}}\,.
\end{align}
We can use these expressions, confirmed fully by numerical simulations, to analyze the behavior of the coupling strength and cooperativity at the different transmission points.\\
\Fref{fig:ScalingCouplingZeta} illustrates how the coupling strength of each mode increases as the single-element reflectivity is increased. We note that two of the three depicted curves are degenerate, in the sense that the curve for $l=4$ (not shown) coincides with that for $l=3$, and that for $l=5$ (not shown) with $l=2$. Similar observations hold for different $N$, where $(N-1)/2$ (if $N$ is even) or $N/2-1$ (if $N$ is odd) curves are twofold degenerate. The order of the respective curves is determined by the parameter $d/L$. This shows that the scaling of the coupling strength with $\zeta$ is essentially the same for all the transmission points.\\
Let us now examine the scaling with the number of elements for a fixed single-element polarizability. The collective coupling strengths, the effective linewidths, and the resulting normalized cooperativities are shown for various transmission points $d=D+d_l$ in \fref{fig:ScalingD0} ($D=0\lambda$) and \fref{fig:ScalingD20} ($D=20\lambda$). When linewidth-narrowing effects are weak, as in \fref{fig:ScalingD0}, the coupling strength of the inner transmission points ($1<l<N-1$) is smaller and saturates more slowly than that of the outer ones ($l=1,N-1$) as $N$ increases; see \fref{fig:ScalingD0}(a). This is due to the fact that the effective cavity length is smaller for these modes, as confirmed by \fref{fig:ScalingD0}(b). When linewidth-narrowing effects are more pronounced, \fref{fig:ScalingD20}, this situation can be reversed:\ the coupling can be stronger for the inner transmission points than for the outer ones. The normalized cooperativity, however, remains largest for the outer transmission points in most such cases.\\
One notable feature in these curves is a local maximum, occurring when $N=2l$. A close inspection reveals that these modes are precisely the (only) ones where the direction of motion of the elements alternates, as shown explicitly in the profile for $N=6$ and $l=3$ in \fref{fig:Profiles}. Mathematically, the feature that gives rise to this anomalous coupling strength is analogous to that pointed out in \eref{eq:NormalisationAnomaly} for $l=1$. Our investigation therefore reveals that the excitation of such modes produces a stronger effect on the cavity field than modes with similar $N$ but otherwise identical parameters. A larger spacing between pairs of elements also affects the effective linewidth of the cavity [see \frefs{fig:ScalingD0}(b) and \ref{fig:ScalingD20}(b)]; the larger the factor $d/L$, the stronger the linewidth-narrowing effect is. Finally, \frefs{fig:ScalingD0}(c) and~\ref{fig:ScalingD20}(c) put these two factors together to show the single-photon cooperativity, normalized to that of a single element inside the same cavity, for the same group of transmission points. We note, in particular, that the abnormally large coupling strengths when $N=2l$ are reflected in the larger cooperativities obtained at these points.

\section{Resilience to imperfections and absorption}\label{sec:Imp}
In this section we address questions regarding the resilience of the mechanism with respect to absorption. The effects of inhomogeneities in the positioning and reflectivity of the individual elements on the achievable couplings and cooperativities were numerically investigated in the Supplemental Information of Ref.~\cite{Xuereb2012c} in the case of the first transmission point. We have checked that a similar sensitivity to deviations from the ideal system is obtained for the other transmission points. The case for absorption, however, is different, since it depends strongly on the transmission point considered.
\par
\begin{figure}[t]%
 \includegraphics[width=\figurewidth]{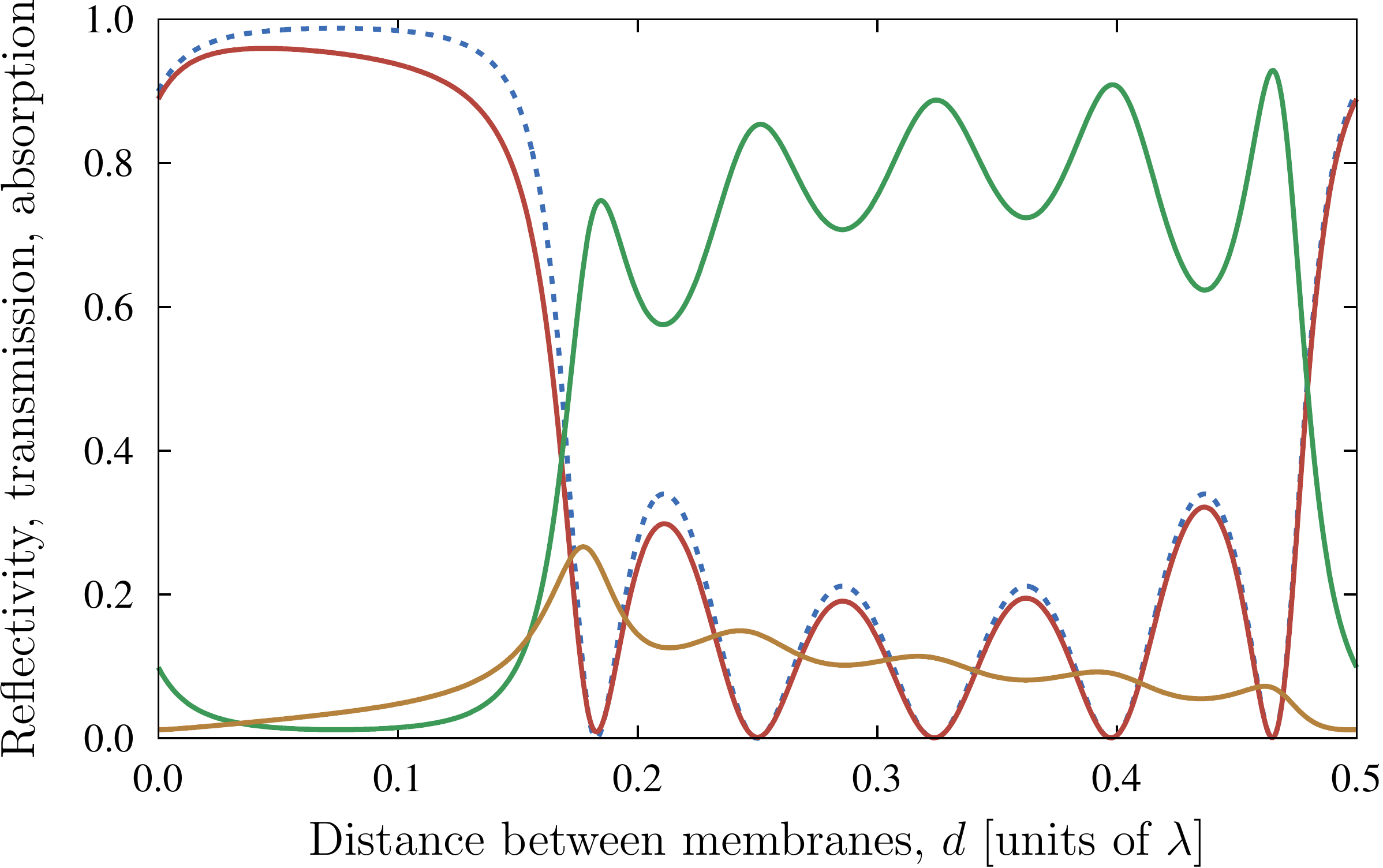}%
\caption{Reflectivity (red), transmission (green), and absorption (orange) for $N=6$ elements with an individual reflectivity of ca.\ $20$\% and an absorption of ca.\ $1.6$\% per element. The dashed blue curve is identical to the solid blue curve in \fref{fig:Reflectivity}. Note that the absorption around the transmissive points is lowest close to $d_{N-1}=d_5$ and highest close to $d_1$.}%
 \label{fig:AbsorptionScan}%
\end{figure}
\begin{figure}[t]%
 \includegraphics[width=\figurewidth]{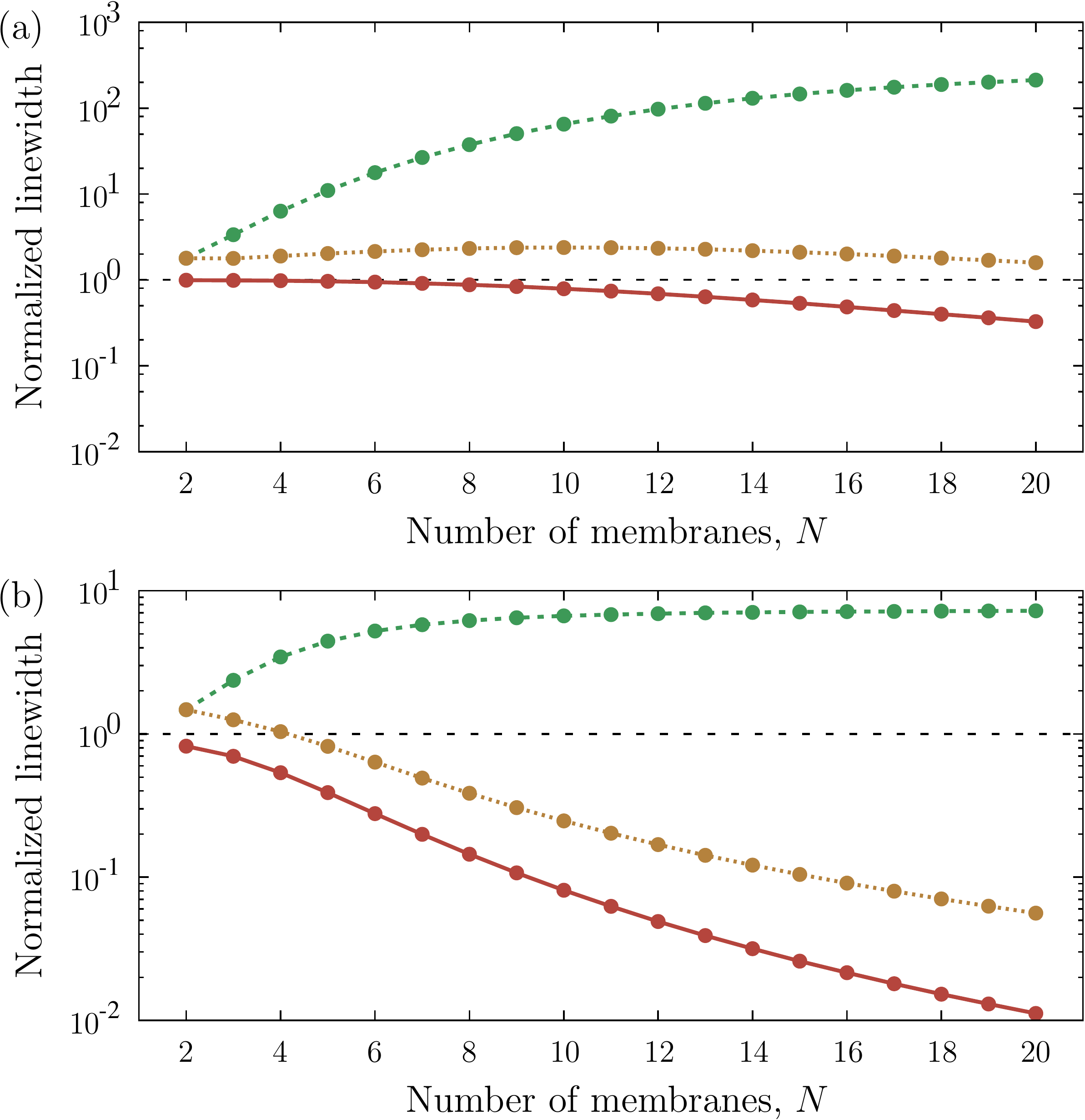}%
\caption{Effect of absorption on the linewidth, shown normalized to the bare-cavity linewidth, in the case of a single-element reflectivity of $99.4$\%. We show $d=D+d_1$ (solid red and dashed green curves), and $d=D+d_{N-1}$ for each $N$ (dotted orange curve), where (a)~$D=0\lambda$ and (b)~$D=20\lambda$. The solid red curve represents non-absorbing scatterers ($\im{\zeta}=0$), and the other two curves $\im{\zeta}=10^{-5}$. The curves for the two values of $d$ coincide in the absence of absorption. Larger inter-element separations make the system more tolerant to higher levels of absorption. The linewidth of the bare cavity is represented by the horizontal dashed black line. (Bare-cavity finesse $\approx\,3\times10^4$, other parameters as in \fref{fig:LineCoop}.)}
 \label{fig:Absorption}%
\end{figure}

%
For a nonzero per-element absorption, the largest amount of absorption appears close to the points where the ensemble is transparent. This is shown in \fref{fig:AbsorptionScan}, where we plot the reflectivity, transmission, and absorption of an ensemble of $6$ elements as the spacing between the elements is scanned; this figure is meant to complement \fref{fig:Reflectivity}. A general feature is that the absorption is largest at $d_{N-1}$ and rather smaller at $d_1$.\\
A systematic study of the effect of absorption on the cavity linewidth is shown in \fref{fig:Absorption}. The cavity linewidth, both in the presence of absorption and in its absence, is calculated numerically by scanning over, and fitting a Lorentzian to, the cavity resonance. An approximate expression for the linewidth in the presence of absorption can be given by taking into account the optical losses in the ensemble through a \emph{small} nonzero $\im{\zeta}$, yielding
\begin{align}
\kappa_\mathrm{eff,abs}^{(l)}&=\frac{c}{2L_\mathrm{eff}^{(l)}}\Biggl(\frac{1}{\abs{Z}\sqrt{Z^2+1}}+2A_l\Biggr)\nonumber\\
&=\kappa_\mathrm{eff}^{(l)}\bigl(1+2A_l\abs{Z}\sqrt{Z^2+1}\bigr)\,,
\end{align}
where the factor $A_l$ corresponds to the single-pass absorption for the ensemble around the working point considered, and $\kappa_\mathrm{eff}$ is evaluated with $\zeta\to-\abs{\zeta}$. For $d=d_1$, $\im{\zeta}\lll1$, $N\geq2$, and $l=1$, $A_1$ is given approximately by
\begin{align}
&2\im{\zeta}\sin(\nu)\bigl[\zeta\cos(\nu)+\sin(\nu)\bigr]U_{N-1}^\prime\bigl[\cos(\pi/N)\bigr]\nonumber\\
&\ \approx\frac{2N\im{\zeta}}{1-\cos\bigl(\frac{2\pi}{N}\bigr)}\Biggl(\sqrt{1+\abs{\zeta}^2}\sin\Biggl\{2\Biggl[\arccos\Biggl(\frac{\cos\bigl(\frac{\pi}{N}\bigr)}{\sqrt{1+\abs{\zeta}^2}}\Biggr)\nonumber\\
&\phantom{\!\!\frac{2N\im{\zeta}}{1-\cos\bigl(\frac{2\pi}{N}\bigr)}\Biggl(}-\arctan(\abs{\zeta})\Biggr]-\arccot(\abs{\zeta})\Biggr\}+1\Biggr)\,,
\end{align}
where $\nu=kd_1$ and $U_n^\prime(x)$ is the first derivative of the $n$\textsuperscript{th} Chebyshev polynomial with respect to its argument. Upon substituting this expression for $A_1$, $\kappa_\mathrm{eff,abs}^{(1)}$ agrees with the corresponding numerically-calculated data shown in \fref{fig:Absorption}. This figure shows explicitly that the effect of absorption decreases for larger inter-element separation, and is stronger for $l=N-1$ than for $l=1$. Indeed, as hinted at by \fref{fig:AbsorptionScan}, the effect of absorption on the linewidth increases monotonically with $l$; for $l\neq1,N-1$, the respective curve lies in the envelope created by dashed green and dotted orange curves in \fref{fig:Absorption}. For large $N$ and $\abs{\zeta}$ (but $\im{\zeta}\lll1$) we find $A_1\approx N\im{\zeta}$.
\par
For $l\neq1$ we could not derive analytical expressions for $A_l$, but our numerical data is fully consistent with the approximate expression
\begin{multline}
\frac{2N\im{\zeta}}{1-\cos\bigl(\frac{2l\pi}{N}\bigr)}\Biggl(\sqrt{1+\abs{\zeta}^2}\sin\Biggl\{2\Biggl[\arccos\Biggl(\frac{\cos\bigl(\frac{l\pi}{N}\bigr)}{\sqrt{1+\abs{\zeta}^2}}\Biggr)\\
-\arctan(\abs{\zeta})\Biggr]-\arccot(\abs{\zeta})\Biggr\}+1\Biggr)\,.
\end{multline}\\
To stay within the frame of the 1D model considered here, a small misalignment in the individual elements can be modeled similarly to absorption, since both effects represent a loss channel for the cavity field. Other detrimental effects of absorption, such as heating, are mitigated by the large coupling strengths obtained, which allow much smaller photon numbers to be used [$(g_\mathrm{sin}^{(1)})^2\varpropto N^3$ increases faster than the absorbed power as $N$ increases]. We note also that at large input powers it might be possible to exploit photothermal forces to further enhance, or change the nature of, the collective optomechanical interaction~\cite{Pinard2008,Restrepo2011,Metzger2004,Usami2012,Xuereb2012b}.

\section{Concluding remarks}
Transmissive optomechanics presents a departure from traditional optomechanical systems in that the reflectivity of a compound element is purposely engineered to be as close to zero as possible. A wealth of interesting effects exist in this regime, not least (i)~the possibility of strongly increasing the optomechanical cooperativity and obtaining strong coupling between a single photon and a single phonon, (ii)~the existence of a linewidth-narrowing mechanism that renders the resolved-sideband regime of optomechanics more accessible, (iii)~the existence of long-range interactions within optomechanical arrays~\cite{Xuereb2012c}, and (iv)~the possibility to enhance optomechanical nonlinearities~\cite{Genes2013}. The system we described may be composed of any periodic array of linearly-interacting polarizable scatterers, e.g, an ensemble of macroscopic dielectric scatterers, or even atoms in an optical lattice, and therefore presents a widely-configurable and robust basis for investigating strong and collective effects in optomechanical or electromechanical setups.

\section*{Acknowledgements}
We acknowledge support from the Royal Commission for the Exhibition of 1851 (A.X.), Austrian Science Fund (FWF):\ P24968-N27 and an STSM grant from the COST Action MP1006 ``Fundamental problems in quantum physics'' (C.G.), and the EU CCQED and PICC projects, and Danish Council for Independent Research under the Sapere Aude program (A.D.). We would also like to thank J.\ Bateman, K.\ Hammerer, I.\ D.\ Leroux, M.\ Paternostro, and H.\ Ritsch for fruitful discussions.

\end{document}